\documentclass[showpacs,twocolumn,prhrenumbers,amsmath,amssymb]{revtex4}
\usepackage{amsmath,amsfonts,latexsym,amssymb,graphicx,graphics,epsfig,subfigure,color,makeidx}
\usepackage{xcolor,epstopdf}
\usepackage{float}
\usepackage{multirow}
\usepackage[colorlinks=true,citecolor=blue,linkcolor=red,anchorcolor=green,urlcolor=cyan]{hyperref}
\usepackage[title]{appendix}
\newcommand {\nn}{\nonumber}

\newcommand {\be}{\begin{equation}}
\newcommand {\ee}{\end{equation}}
\newcommand {\beq}{\begin{eqnarray}}
\newcommand {\eeq}{\end{eqnarray}}

\graphicspath{{plots/}}
\makeatletter

\begin{document}
\title{Imprints of the Lorentz-symmetry breaking on the precessing jet nozzle of M87*}

\author{Tao-Tao Sui$^{a}$\footnote{suitaotao@aust.edu.cn}}
\author{Xiang-Cheng Meng$^{b}$\footnote{mxiangcheng2023@lzu.edu.cn}}
\author{Xin-Yang Wang$^{a}$\footnote{wangxy@aust.edu.cn, Corresponding author}}

\affiliation{$^{a}$Center for Fundamental Physics, School of Mechanics and Photoelectric Physics, Anhui University of Science and Technology, Huainan, Anhui 232001,  China\\
$^{b}$Institute of Theoretical Physics, Research Center of Gravitation,
and School of Physical Science and Technology, Lanzhou University, Lanzhou 730000, People’s Republic of China}

\begin{abstract}
{The approximately 11-year jet precession period observed in M87* strongly suggests that the supermassive rotating black hole with a tilted accretion disk, which could provide a powerful constraint for confining the parameters of black hole. In this paper, our aim is to utilize the observations of M87* to preliminarily constrain the parameters of the rotating black hole in Bumblebee gravity by modeling the motion of the tilted accretion disk particle with the spherical orbits. We compute spherical orbits and ISSOs, demonstrating that the conserved quantities energy $\mathcal{E}$, angular momentum $\mathcal{L}$, and Carter constan $\mathcal{K}$ depend on $(r,a,\ell,\zeta)$, exhibiting distinct behaviors for prograde and retrograde orbits. For prograde orbits, the ISSO radius $r_{ISSO}$ decreases with spin parameter $a$ and LSB parameter $\ell$ and increases with the tilt angular $\zeta$, whereas the opposite trends occur for retrograde orbits. Angular analysis shows that $\theta$ oscillates within $(\pi/2-\zeta, \pi/2+\zeta)$, while $\phi$ increases approximately linearly, enabling the determination of the oscillation period $T_\theta$, azimuthal accumulation $\phi(T_\theta)/\pi$, and precession angular velocity $\omega_t$. Using the observed jet precession period $T=11.24 \pm 0.47$ years with a fixed tilt $\zeta=1.25^\circ$, the warp radius $r/M$ ranges from $(5.73,25.15)$ for prograde and $(6.16,26.46)$ for retrograde orbits, increasing with $a$ or $\ell$. Comparisons with Kerr limits ($r/M=14.12$ prograde, $16.1$ retrograde) suggest that $r/M>16$ may indicate a non-vacuum Bumblebee vector field. Incorporating the EHT shadow $\theta_{sh}=42\pm3\mu$as further constrains $r/M$ to $(5.82,22.61)$ and $(6.17,24.74)$, with discrepancies of $0.05$–$1.96$. These results highlight the dominant role of $a$ and establish a framework for constraining black hole properties within alternative gravity theories.

}
\end{abstract}

\maketitle

\section{Introduction}\label{firstpart}
Lorentz invariance, a cornerstone of both general relativity and quantum field theory, may exhibit subtle violations that point toward deeper underlying physical principles. Evidence from unified canonical theories and observations of ultra–high-energy cosmic rays \cite{Takeda:1998ps} suggests that spontaneous Lorentz-symmetry breaking (LSB) could occur at high energy scales. Empirically, Lorentz-violation effects are typically detectable only at low energies, where they manifest as ``relic signatures’’ of quantum-gravity phenomena. These effects can also be systematically described within the framework of effective field theory \cite{Kalb:1974yc,Kostelecky:1994rn,Colladay:1998fq,Devecioglu:2024uyi}.

Kostelecky and Samuel introduced a simple yet influential model of spontaneous LSB, now commonly known as bumblebee gravity \cite{Kostelecky:1989jw}. In this framework, the coupling between a bumblebee vector field $B_\mu$, which acquires a nonzero vacuum expectation value (VEV), and the gravitational field leads to a background spacetime that no longer preserves full Lorentz symmetry, thereby yielding an incomplete symmetry structure \cite{Kostelecky:2000mm,Kostelecky:2002ca,Bertolami:2003qs,Cambiaso:2012vb}. Owing to its conceptual simplicity modifying gravitational dynamics through nonminimal coupling with a spontaneously Lorentz-violating field, and its broad physical implications, bumblebee black hole solutions and Lorentz-violating effects have attracted considerable interest in recent years \cite{Bluhm:2004ep,Bailey:2006fd,Bluhm:2008yt,Seifert:2009gi,Maluf:2014dpa,Assuncao:2019azw,Escobar:2017fdi}.

Then, many study have constructed a variety of black hole solutions within the bumblebee gravity framework, including Schwarzschild-like static spherically symmetric black holes \cite{Bertolami:2005bh, Casana:2017jkc}, global monopole configurations \cite{Gullu:2020qzu}, solutions incorporating cosmological constants \cite{Capelo:2015ipa, Maluf:2020kgf, Liu:2024axg}, and spherically symmetric models involving Einstein–Gauss–Bonnet corrections \cite{Ding:2021iwv}, among others. Bumblebee gravity reveals a range of novel physical phenomena and significantly broadens our understanding of the behavior of compact objects in modified gravity theories, an endeavor central to advances in modern theoretical physics \cite{Liu:2022dcn, Xu:2023xqh, Zhang:2023wwk, Chen:2020qyp, Chen:2023cjd, EslamPanah:2025zcm, Liu:2024wpa, AraujoFilho:2024ctw, Liu:2025bpp, AraujoFilho:2025fwd, Shi:2025plr, Guo:2023nkd, Mai:2023ggs, Mai:2024lgk}.

It is noteworthy that real astrophysical black holes generally possess angular momentum; consequently, realistic observational modeling requires rotating black hole solutions. In this context, rotating black hole solutions within the bumblebee gravity framework were subsequently constructed in Ref.~\cite{Ding:2019mal}, where the influence of the LSB parameter on the resulting black hole shadow was analyzed. Building on this foundation, extensive studies have examined the strong-field gravitational lensing properties \cite{Kuang:2022xjp}, shadow characteristics \cite{Wang:2021irh}, accretion disk structure \cite{Liu:2019mls}, superradiant instabilities \cite{Jiang:2021whw}, and particle dynamics \cite{Li:2020dln} associated with this rotating solution. Moreover, stringent constraints on the LSB parameter introduced by the bumblebee field have been derived using a range of astrophysical observations \cite{Wang:2021gtd, Gu:2022grg}. These studies provide valuable tests of the bumblebee model and offer potential avenues for identifying LSB signatures in the vector field.

On the other hand, the Event Horizon Telescope (EHT) Collaboration’s first image of the supermassive black hole M87* provided direct observational confirmation of the black hole shadow, transforming a long-standing theoretical prediction into an empirical reality \cite{EventHorizonTelescope:2019dse,EventHorizonTelescope:2019uob,EventHorizonTelescope:2019jan}. The measured shadow, characterized by a deviation from circularity of $\Delta C \lesssim 0.1$, an axis ratio of $1 < D_x \lesssim 4/3$, and an angular shadow diameter $\psi_d=3\sqrt{3}(1\pm0.17)\psi_g$, where the angular gravitational radius is $\psi_g=3.8\pm0.4\mu$as, has been effectively employed to constrain Kerr-like black hole models \cite{Meng:2022kjs,Kuang:2022ojj,Xavier:2023exm,Sui:2023rfh}. In addition, the properties of black holes and other exotic compact objects can be probed through gravitational-wave observations \cite{Destounis:2020kss,Destounis:2021mqv,Destounis:2023gpw}, and ongoing efforts aim to constrain non-vacuum black holes situated in astrophysical environments \cite{Cardoso:2021wlq,Cardoso:2022whc,Destounis:2022obl,Figueiredo:2023gas}.
 
In addition to accreting matter, the supermassive black hole at the center of M87 launches bright relativistic jets that exhibit significant angular displacement near the black hole \cite{Junor:1999Natur,Hada:2011Natur,Lu:2023bbn}. Analysis of 22 years of radio observations revealed a periodic oscillation in the jet’s position angle \cite{Cui:2023uyb}, which supported by general relativistic magnetohydrodynamic simulations, was attributed to Lense–Thirring precession driven by a tilted accretion disk. The inferred precession cone half-opening angle is $1.25^\circ \pm 0.18^\circ$, with a period of $11.24 \pm 0.47$ years and an angular velocity of $0.56 \pm 0.02$ radians per year, offering strong evidence for a tilted disk whose structure evolves with radius: the inner region, extending from the innermost stable spherical orbits (ISCO), aligns with the black hole’s equatorial plane through the Bardeen–Petterson mechanism \cite{Bardeen:1975ApJ}, while beyond the warp radius the inclination gradually increases. Tilted disks are common in various astrophysical systems \cite{Begelman:2006bi,Wijers:1998wx,Chiang:2003gi,Casassus:2015,Lodato:2013,Martin:2008ib}, and simplified models of their dynamics, such as analyses of spherical orbit precession \cite{AlZahrani:2023xix} can elucidate quasi-periodic oscillations and constrain black hole parameters via jet precession \cite{Teo:2020sey,Kopacek:2024pfd,Rana:2019bsn,Iorio:2024eey}, while also explaining observed changes in disk scale and non-collinear jet morphology \cite{Cui:2024ggx}.

In Ref. \cite{Islam:2024sph}, the authors derived a slightly modified rotating black hole metric in Bumblebee gravity, which differs marginally from previous results \cite{Ding:2019mal}. By combining these theoretical findings with EHT observations of the M87* black hole shadow, they imposed tighter constraints on the LSB parameter and highlighted subtle effects of the Lorentz-violating field. Similarly, Wei et al. \cite{Wei:2024cti} employed M87 jet observations to establish a correlation between the black hole spin parameters and the twist radius of a tilted accretion disk, suggesting that the precession of black hole jet nozzles provides a promising probe of the strong-gravity regime around supermassive black holes. Motivated by these studies, we will investigate the dynamics of massive particles around a rotating black hole in Bumblebee gravity, and, using the parameter ranges derived in Ref. \cite{Islam:2024sph} from M87* shadow and jet observations, we determine the corresponding twist-radius range of the tilted accretion disk. In our analysis, we adopt a simplified model of the black hole environment, which allows us to constrain relevant black hole parameters, enhance our understanding of fundamental deviations in astrophysical environments, and refine constraints on quantum gravity. This approach may provide deeper insights into the structure of spacetime at the Planck scale and the underlying architecture of the universe.

The remainder of the paper is organized as follows. In Section \ref{secondpart}, we briefly review the rotating black hole solution in Bumblebee gravity obtained via the modified Newman–Janis algorithm. Section \ref{thirdpart} examines the motion of massive particles around the rotating black hole with LSB parameter, including spherical orbits and the innermost stable spherical orbits. In Section \ref{secPOSO}, we study the precession of spherical orbits and analyze the influence of black hole parameters on this precession. The constraints on the parameters of the rotating black hole in Bumblebee gravity derived from the precessing jet nozzle and the angular radius of the M87* shadow are shown in Section \ref{secConstrain}. Finally, Section \ref{conclusion} summarizes our main results and provides concluding remarks.

\section{Black Holes in Bumblebee Gravity}\label{secondpart}
In this section, we will give a succinct review of black hole in Bumblebee gravity, which can cause spontaneous Lorentz symmetry breaking by introducing a non-vanishing vacuum vector field \citep{Kostelecky:1989jw,Bluhm:2004ep}. The corresponding action of bumblebee gravity described with \citep{Casana:2017jkc,Ding:2019mal}:
\begin{align}\label{eqn:100}
S=\int d^4x\sqrt{-g}&\Big[\frac{1}{16\pi}\big(R+\varrho B^\mu B^\nu R_{\mu\nu}\big) \nn\\
&-\frac{1}{4}B^{\mu\nu}B_{\mu\nu}-V(B^\mu)\Big],
\end{align}
where $\varrho$ is a coupling constant, and the bumblebee field gets a non-vanishing vacuum expectation value $\langle B^\mu \rangle = b^\mu$. $B_{\mu\nu}$ is the bumblebee field strength $B_{\mu\nu}=\partial_{\mu}B_{\nu}-\partial_{\nu}B_{\mu}$, and $V(B^\mu)$ is the  potential of the bumblebee vector field with $V=V(B^{\mu}B_{\mu}\pm b^2)$. 

\begin{figure}[!htbp]
\includegraphics[width=0.35\textwidth]{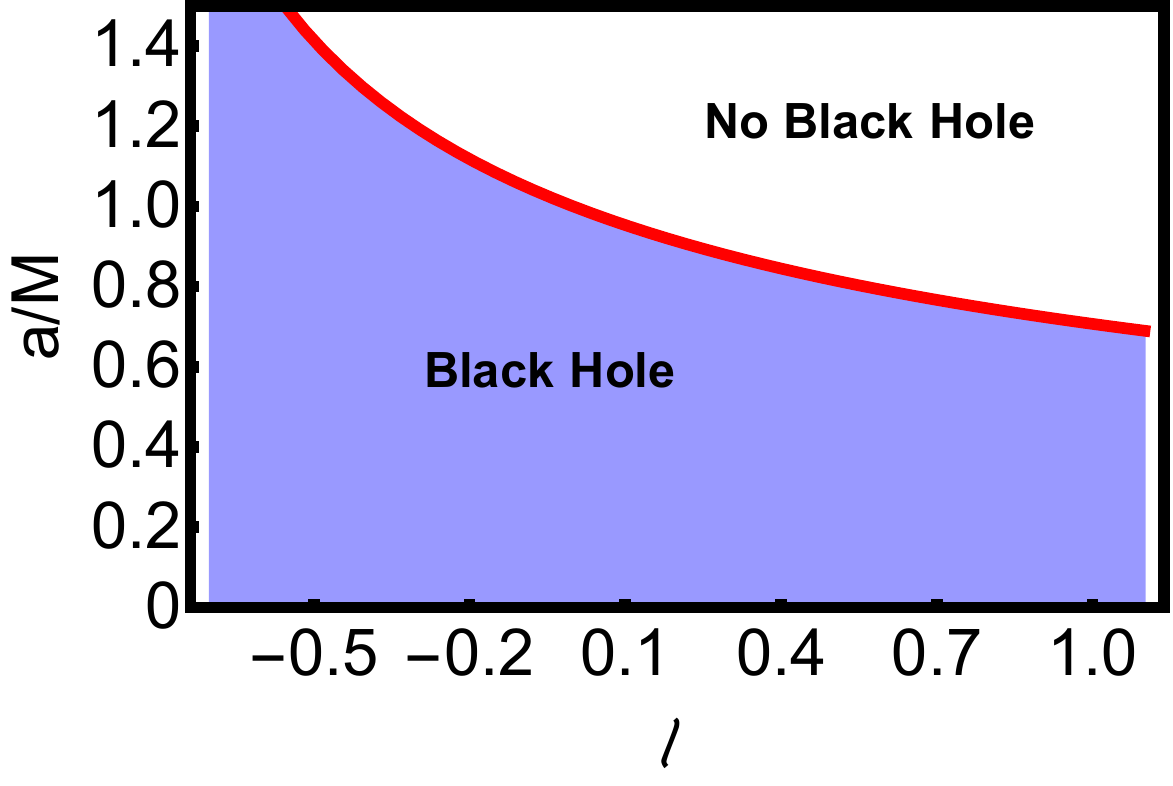}
\caption{Parameters space of spin and LSB $(a,\ell)$. The colored region represents the allowed range of black holes.}\label{event condtion}
\end{figure}

By combining with the modified Newman-Janis algorithm \cite{Azreg-Ainou:2014pra,Azreg-Ainou:2014aqa}, we can obtain the rotating spacetime of Bumblebee gravity with \cite{Islam:2024sph}
\begin{align}\label{metric3}
ds^2=&-\left(1-\frac{2M(r)r }{\Sigma}\right)dt^2+\frac{\Sigma}{\Delta}dr^2+\Sigma d\theta^2\nn\\
     &+\frac{4aM(r)r}{\Sigma}\sin^2\theta dtd\phi-\frac{\mathcal{A}\sin^2\theta~}{\Sigma} d\phi^2,
\end{align}
where
\begin{align}
&\Delta=r^2+a^2 -2 M(r)r,~~\mathcal{A} = (r^2+a^2)^2-a^2 \Delta \sin^2\theta,\nn \\
&M(r)=\frac{M(1+\frac{r\ell}{2M})}{1+\ell},~~\Sigma = r^2 +a^2\cos^2\theta.
\end{align}
Here, the parameters ($M, a, \ell=\varrho b^2$) are the mass, spin and LSB parameter of black hole, respectively. For $\ell = 0$, the solution reduces to a standard spherical black hole \citep{Casana:2017jkc,Ding:2019mal}. The radial coordinate of the event horizon can be determined with the equation $g^{rr}=\Delta=0$, and with the expression $r_{\rm h}=M + \sqrt{M^2-a^2(1+\ell)}$, which imposes the constraint $|a|\leq\frac{M}{\sqrt{1+\ell}}$. Notably, for $\ell<0$, the allowed range of the spin parameter $a$ can exceed $M$, and Fig. \ref{event condtion} illustrates the parameter regions corresponding to black hole solutions (colored areas), while the uncolored regions correspond to configurations without horizons. For a comprehensive discussion of the properties of rotating black holes in Bumblebee gravity, we refer the reader to Refs. \cite{Islam:2024sph,Ding:2019mal}.

\section{Motion of massive particles around Black Holes in Bumblebee Gravity}\label{thirdpart}

In this section, we will investigate the kinematic properties of massive test particles around the rotating Bumblebee black hole. For a given geometry, the geodesics of massive test particles with unit mass can be governed by the Hamilton Jacobi equation with
\begin{align}\label{eqHJeq}
\frac{\partial S}{\partial\tau}=-\frac{1}{2}g^{\mu\nu}\frac{\partial S}{\partial x^{\mu}}\frac{\partial S}{\partial x^{\nu}},
\end{align}
where $\tau$ is an affine parameter along the geodesics and $S$ is the Jacobi action. The separable form of $S$ can be written as
\begin{align}\label{eqJacobiAction}
S=\frac{1}{2}\tau-\mathcal{E}t+\mathcal{L}\phi+S_{r}(r)+S_{\theta}(\theta),
\end{align}
where the parameters $\mathcal{E},\mathcal{L}$ are the energy and the angular momentum of the test particle, which are associating with the Killing fields $\partial_t$ and $\partial_\phi$, respectively. By combining the above two equations, we can get the expressions of four equations of motion (EOM) in the directions $(t, r, \theta, \phi)$ as:
\begin{align}
\rho^{2}\frac{dt}{d\tau}=&aP_{\theta}+\frac{(r^{2}+a^{2})P_{r}}{\Delta},\\
~~\rho^{2}\frac{d\phi}{d\tau}=&\frac{P_{\theta}}{\sin\theta^2}+\frac{aP_{r}}{\Delta}\label{eqttl},\\
\rho^{2}\frac{dr}{d\tau}=&\pm\sqrt{\mathcal{R}(r)}\label{equr},\\
\rho^{2}\frac{d\theta}{d\tau}=&\pm\sqrt{\Theta(\theta)}\label{equtheta},
\end{align}
with
\begin{align}
&P_{\theta}=\mathcal{L}-a\mathcal{E}\sin\theta^2,\nn\\&P_{r}=\mathcal{E}(r^2+a^2)-a\mathcal{L}
\nn\\
&\mathcal{R}(r)=P_{r}^{2}-\Delta\Big(r^{2}+\mathcal{K}+(a\mathcal{E}-\mathcal{L})^2\Big),\label{rdirection}\\
&\Theta(\theta)=\mathcal{K}+(a\mathcal{E}-\mathcal{L})^2-a^2\cos^{2}\theta-\frac{P_{\theta}^2}{\sin\theta^2}\label{thetadirection}.
\end{align}
Here $\mathcal{K}$ is another conserved quantity along each geodesic, called the Carter constant, which is corresponding to the Killing-Yano tensor \cite{Carter:1968ks}. 

For the equatorial orbit, with the condition $\theta=\pi/2,\frac{d\theta}{d\tau}=0$, the Carter constant vanish to zero, $\mathcal{K}=0$. Consequently, for the orbits deviating from the equatorial plane, the Carter constant $\mathcal{K}$ will be a function as $\mathcal{K}=\mathcal{K}(a,\mathcal{E},\mathcal{L},\theta)$. From the specific expression of $\theta$ direction Eq. \eqref{thetadirection}, we can see that the $\theta$-motion of the massive test particle exhibits a symmetry with $\theta=\pi/2$. Therefore, for an off-equatorial orbit, the $\theta$-motion can be  characterized as oscillating in the vicinity of the equatorial plane within the range $(\frac{\pi}{2}-\zeta,\frac{\pi}{2}+\zeta)$, where $\zeta$ can be considered as the tilt angle to the equatorial plane with $\zeta\in \left(0,\frac{\pi}{2}\right)$. Because of the particle turning back at two points $\theta=\frac{\pi}{2}\pm\zeta$, we can get $\frac{d\theta}{d\tau}=0$, and 
\begin{align}\label{eqcaterpar}
\mathcal{K}=\mathcal{L}^{2}\tan^{2}\zeta+a^2(1-\mathcal{E}^2)\sin^{2}\zeta.
\end{align}

\subsection{Spherical orbits}
Recently, Ref. \cite{AlZahrani:2023xix} investigated spherical orbits in the presence of tilted accretion disks and demonstrated that their precession can be analyzed from the perspective of local observers. Wei et al. \cite{Wei:2024cti} subsequently examined the energy $\mathcal{E}$, angular momentum $\mathcal{L}$, and Carter constant $\mathcal{K}$ associated with spherical orbits, while Ref. \cite{Kopacek:2024pfd} provided a detailed analysis of the corresponding quantities for the innermost stable spherical orbits (ISSOs). In the present study, to compute the precession period of spherical orbits, we investigate the influence of the LSB parameter $\ell$ on the energy $\mathcal{E}$ and angular momentum $\mathcal{L}$ under different values of spin parameter $a$ and tilt angle $\zeta$. 

For the spherical orbit, the motion of radial r-direction, Eq. \eqref{equr}  should satisfy the conditions ($\dot{r}=\ddot{r}=0$), which are equivalent to $\mathcal{R}(r)=\mathcal{R}^\prime(r)=0$, with the prime denotes the derivative to $r$. By solving the spherical orbit conditions, we can obtain the expressions of the energy $\mathcal{E}$ and angular momentum $\mathcal{L}$, and for brevity the corresponding explicit forms of $\mathcal{E}$ and $\mathcal{L}$ can be shown with 
\begin{align}\label{sporbit}
\mathcal{E}=\mathcal{E}\left(r,a,\ell,\zeta\right), ~\mathcal{L}=\mathcal{L}\left(r,a,\ell,\zeta\right),
\end{align} 

To facilitate a more precise discussion of spherical orbit properties, we adopt the following conventions: the black hole spin parameter is taken as $a\ge0$; orbits with positive angular momentum $\mathcal{L}>0$ are defined as prograde, while those with negative angular momentum $\mathcal{L}<0$ are defined as retrograde.

\begin{figure}[!htb]
\subfigure[{}]{\includegraphics[width=0.226\textwidth]{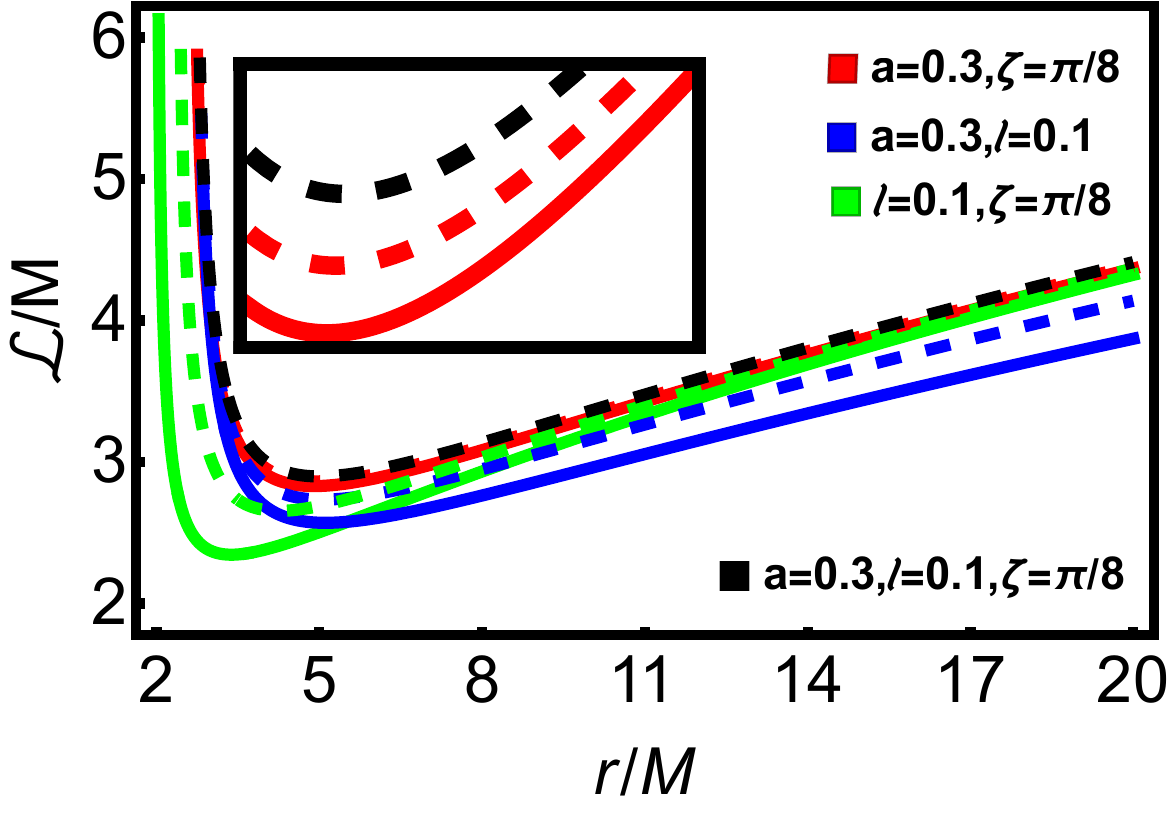}\label{pangular}}
\subfigure[]{\includegraphics[width=0.238\textwidth]{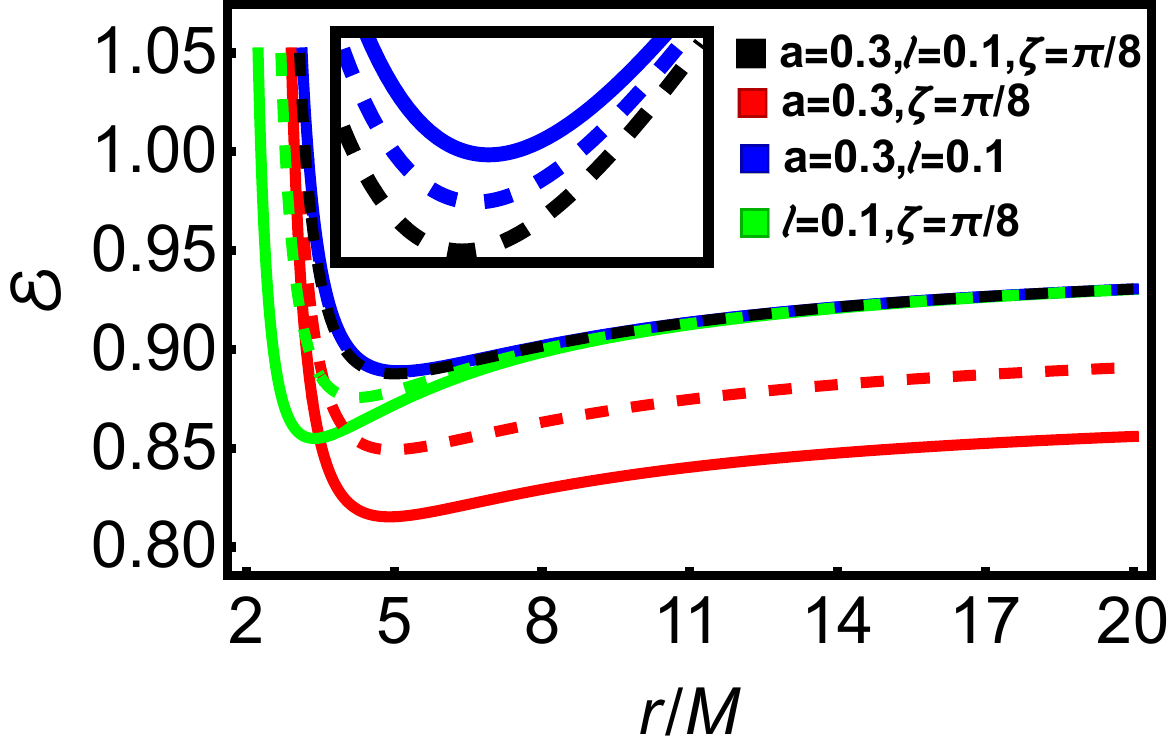}\label{penergy}}
\subfigure[]{\includegraphics[width=0.236\textwidth]{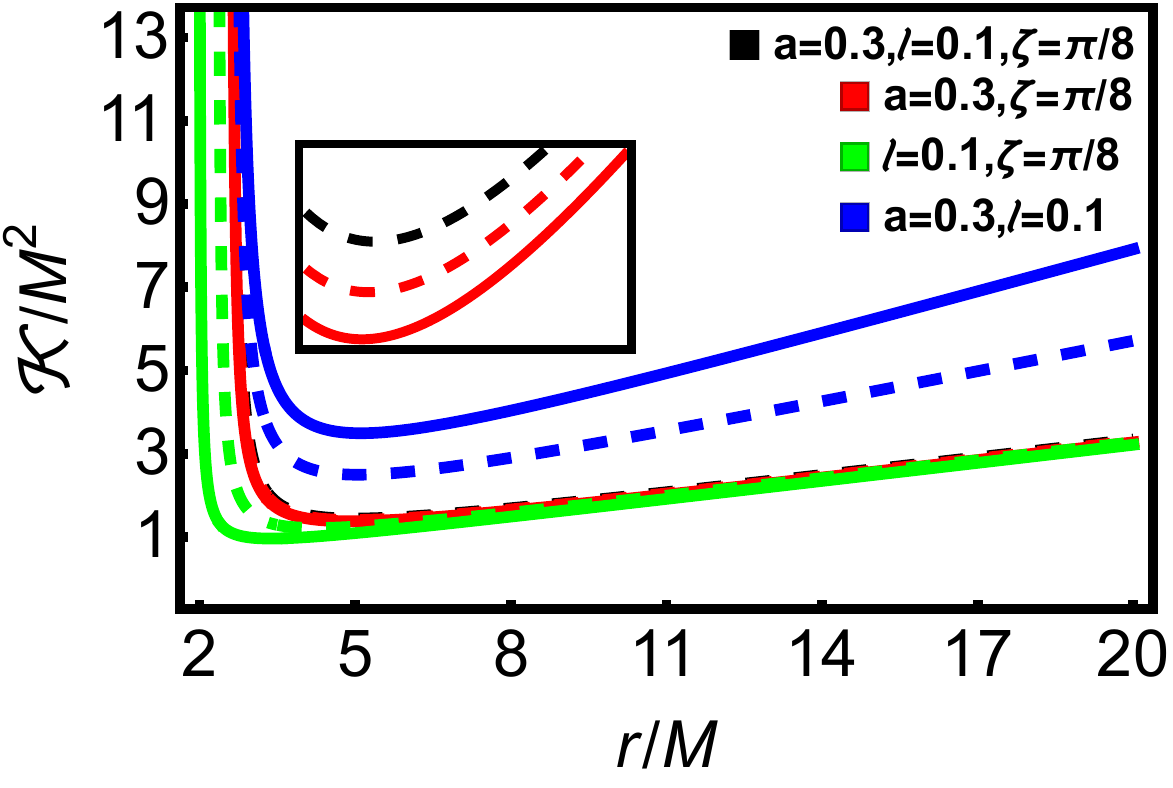}\label{pcarter}}
\subfigure[]{\includegraphics[width=0.226\textwidth]{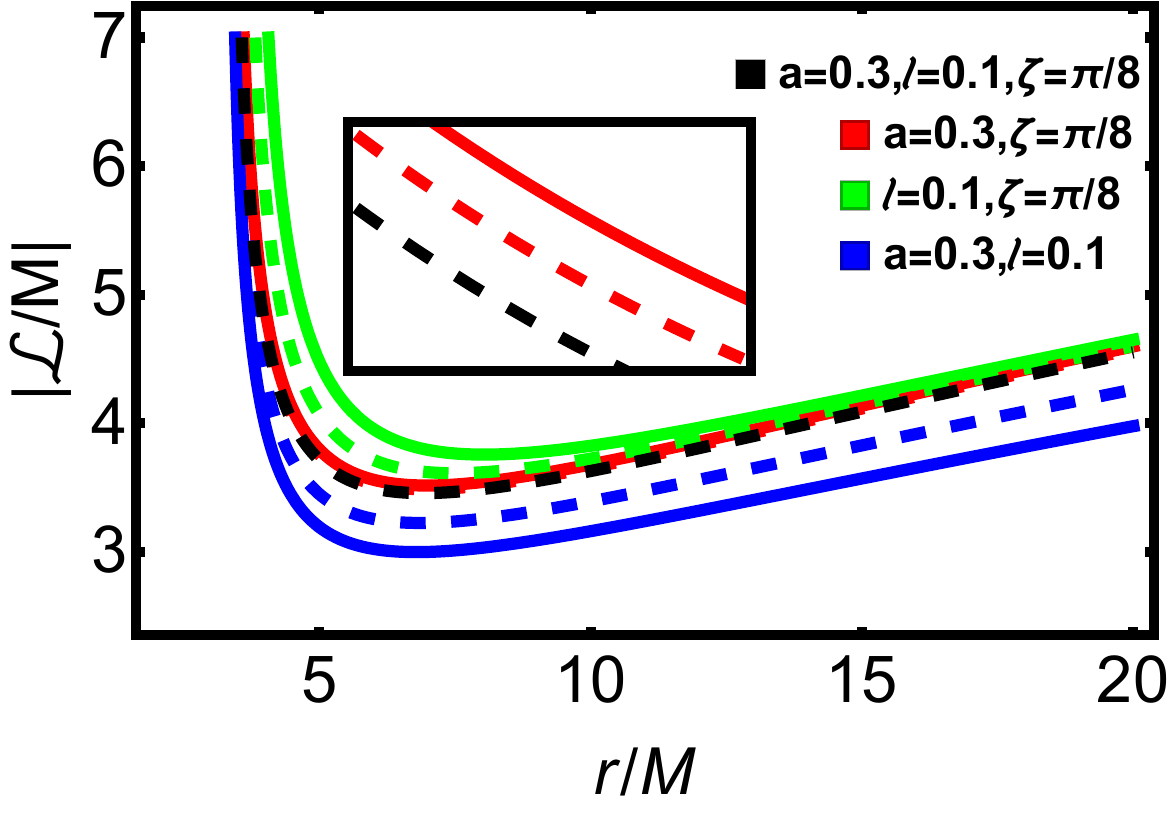}\label{nangular}}
\subfigure[]{\includegraphics[width=0.236\textwidth]{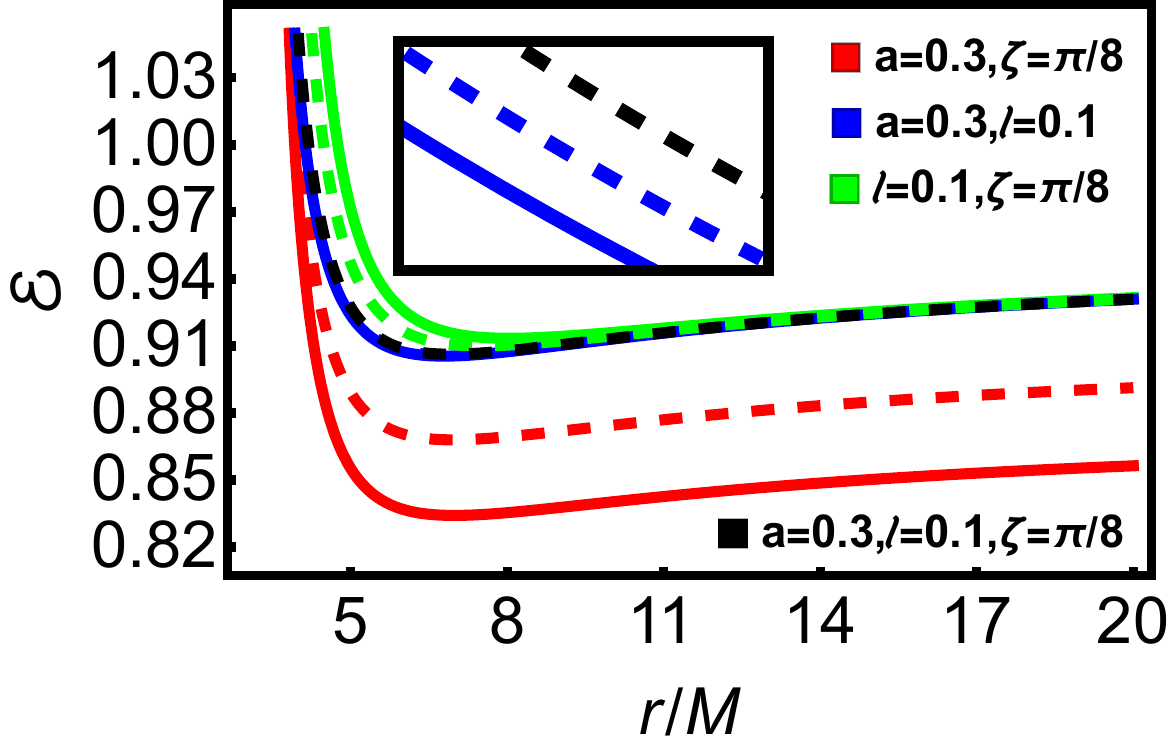}\label{nenergy}}
\subfigure[]{\includegraphics[width=0.23\textwidth]{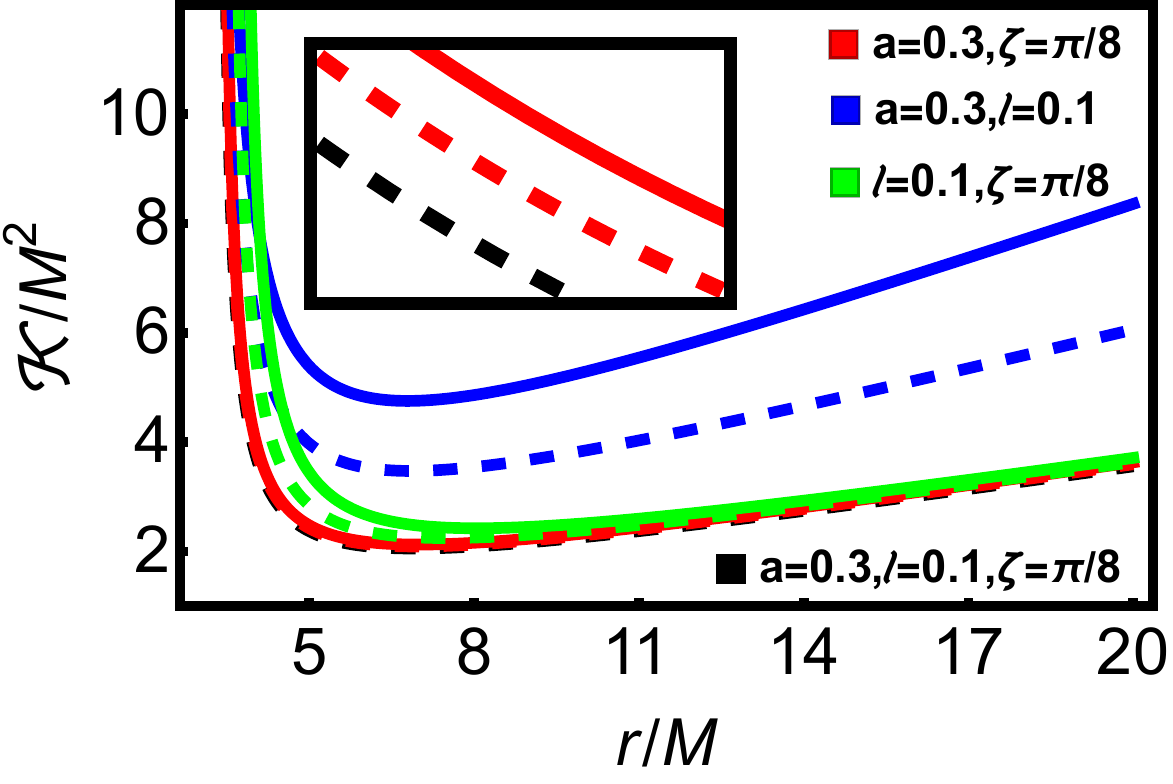}\label{ncarter}}
\caption{Angular momentum $\mathcal{L}$, energy $\mathcal{E}$ and Carter constant $\mathcal{K}$ for the spherical orbits with different parameters. Figures (a)-(c) show the prograde orbits and (d)-(f) describe the retrograde orbits. The dashed and solid red curves correspond to $\ell=0.2~\text{and}~\ell=0.3$, the dashed and solid blue curves correspond to $\zeta=\pi/6~\text{and}~\zeta=\pi/5$, and the dashed and solid green curves correspond to $a=0.5~\text{and}~a=0.7$, respectively.}\label{figle}
\end{figure}

Figure \ref{figle} illustrates the effects of the parameters ($a,\ell,\zeta$) on the properties of spherical orbits, including the angular momentum $\mathcal{L}$, energy $\mathcal{E}$ and Carter constant $\mathcal{K}$. For prograde orbits, increases in both the LSB parameter $\ell$ and the spin parameter $a$ tend to suppress $\mathcal{L}$, $\mathcal{E}$ and $\mathcal{K}$. In contrast, increasing the tilt angle $\zeta$ enhances the energy $\mathcal{E}$ and Carter constant $\mathcal{K}$, while reducing the angular momentum $\mathcal{L}$. For retrograde orbits, the Carter constant $\mathcal{K}$ exhibits a positive correlation with variations in all three parameters ($a,\ell,\zeta$). The numerical values of angular momentum $\mathcal{L}$ increases with the LSB parameter $\ell$ and spin parameter $a$, while, decreases as tilt angle $\zeta$ grows. The energy $\mathcal{E}$ decreases with increasing $\ell$ and $\zeta$, whereas it rises with an increase in spin parameter $a$.

\subsection{Innermost stable spherical orbits}\label{secISSOlCO}
In the previous subsection, we examined the characteristics of spherical orbits under varying parameters. In general, not all spherical orbits are stable, and their stability is determined by the value of $\mathcal{R}^{\prime\prime}(r)$. For the innermost stable spherical orbits (ISSOs), the following conditions must be satisfied:
\begin{align}\label{issocondition}
\mathcal{R}(r)=\mathcal{R}^{\prime}(r)=\mathcal{R}^{\prime\prime}(r)=0.
\end{align}
Besides, we can note that the radius of ISSOs always locate at the extremal points of each curve of the spherical orbits, which are shown in Fig. \ref{figle}, and the corresponding equations to the ISSOs can be shown as
\begin{align}
\Big(\frac{d\mathcal{L}}{dr}\Big)_{a,\ell,\zeta}=\Big(\frac{d\mathcal{E}}{dr}\Big)_{a,\ell,\zeta}=\Big(\frac{d\mathcal{K}}{dr}\Big)_{a,\ell,\zeta}=0.
\end{align}

\begin{figure}[!htb]
\subfigure[]{\includegraphics[width=0.22\textwidth]{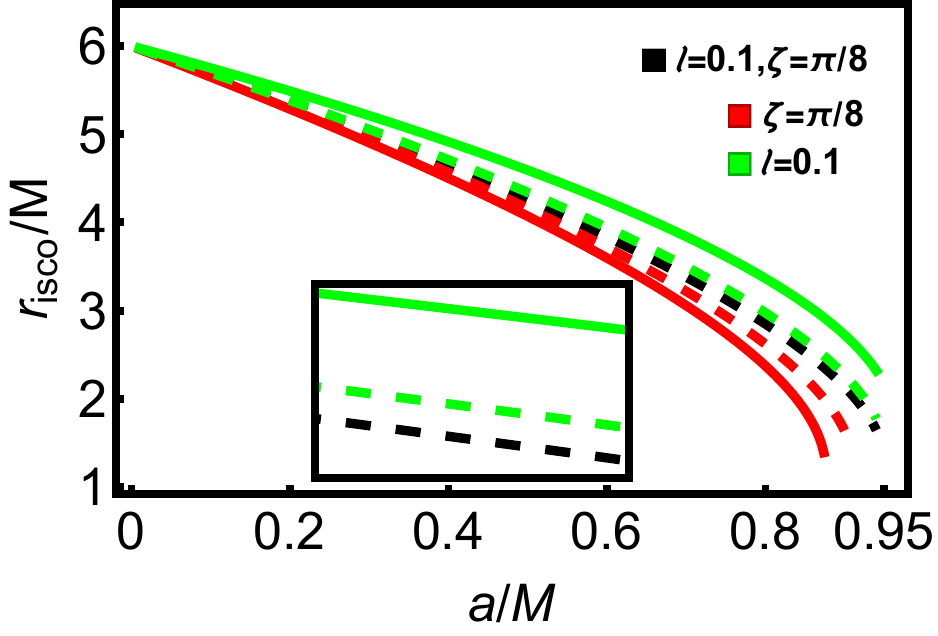}\label{prisco}}
\subfigure[]{\includegraphics[width=0.236\textwidth]{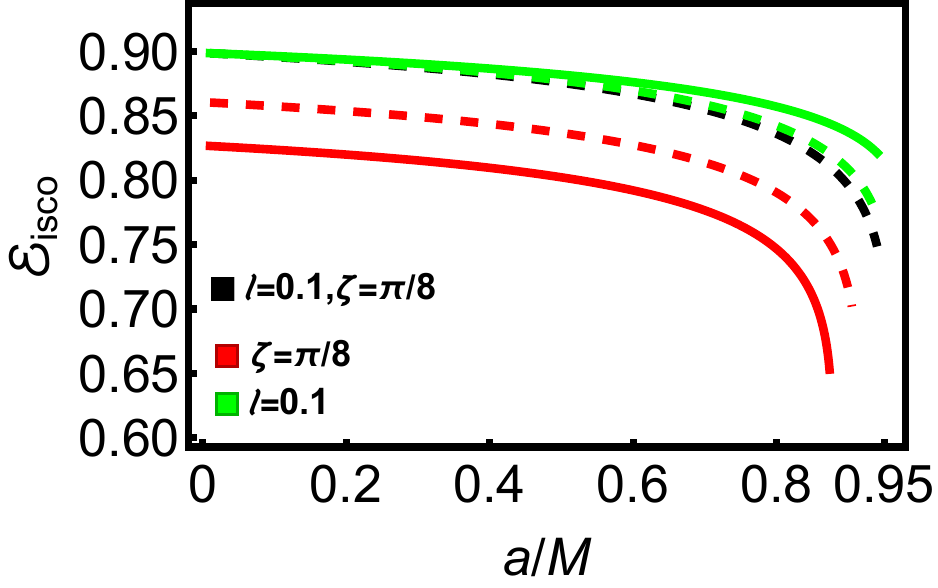}\label{peisco}}
\subfigure[]{\includegraphics[width=0.236\textwidth]{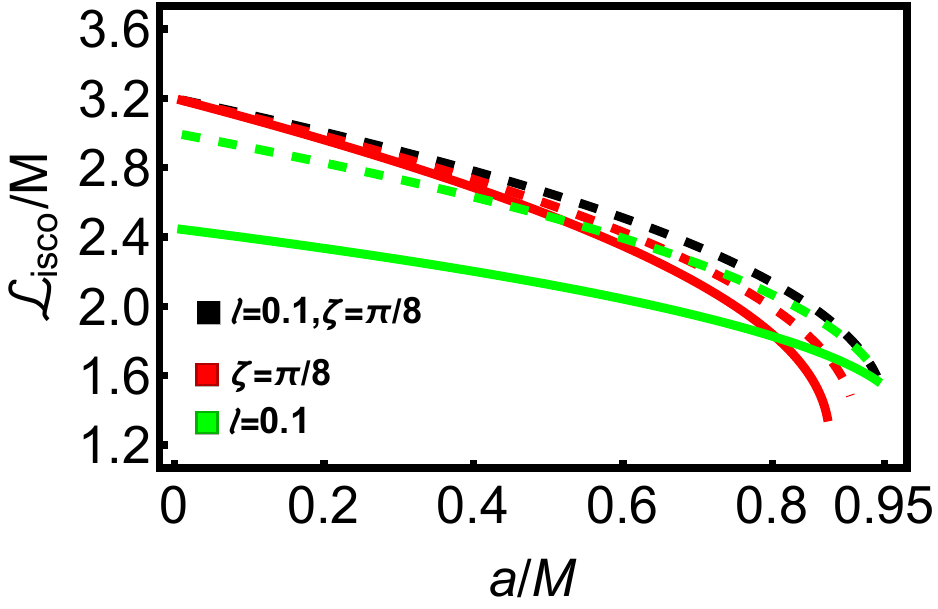}\label{plisco}}
\subfigure[]{\includegraphics[width=0.225\textwidth]{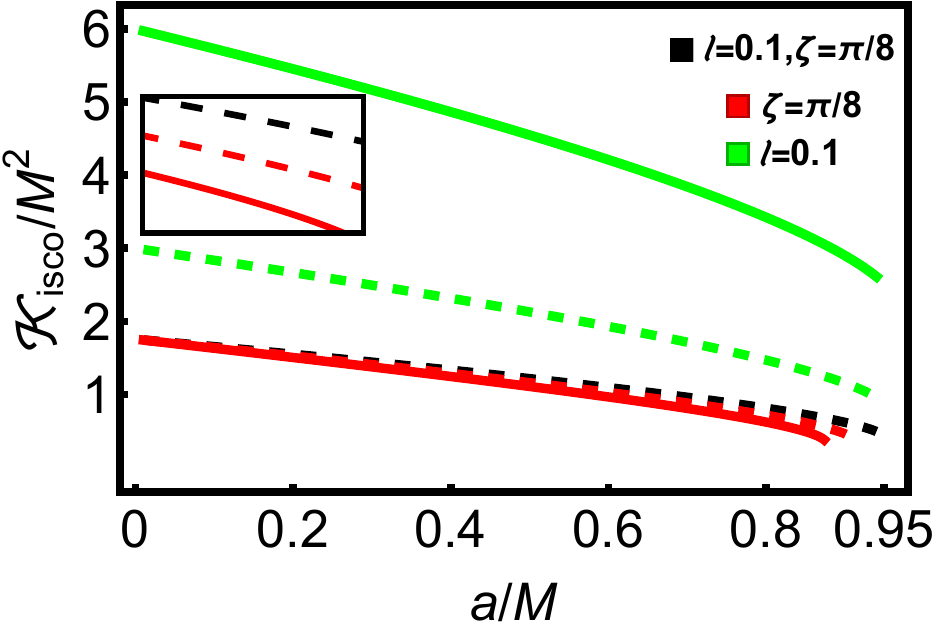}\label{pcisco}}
\subfigure[]{\includegraphics[width=0.236\textwidth]{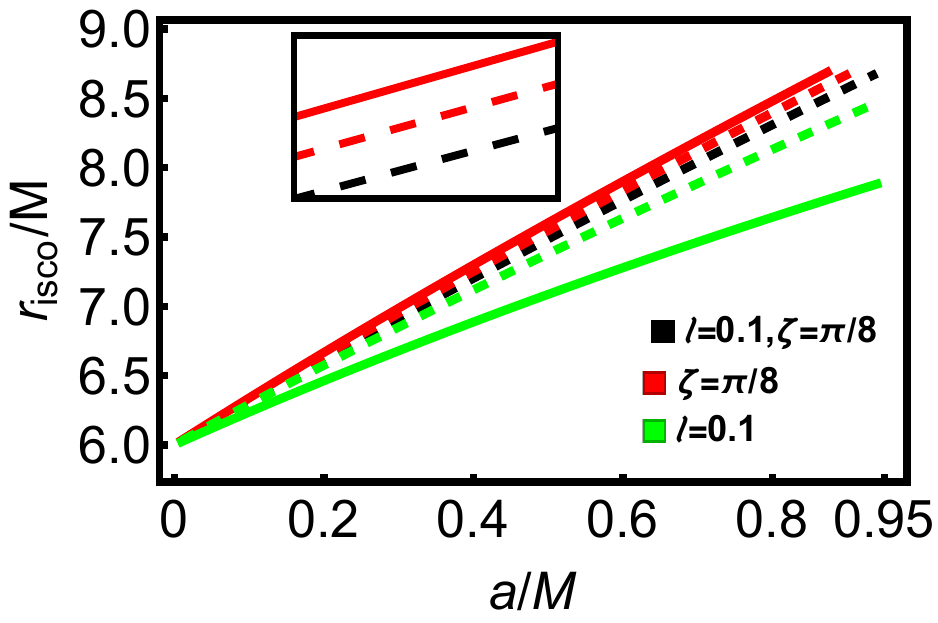}\label{nrisco}}
\subfigure[]{\includegraphics[width=0.236\textwidth]{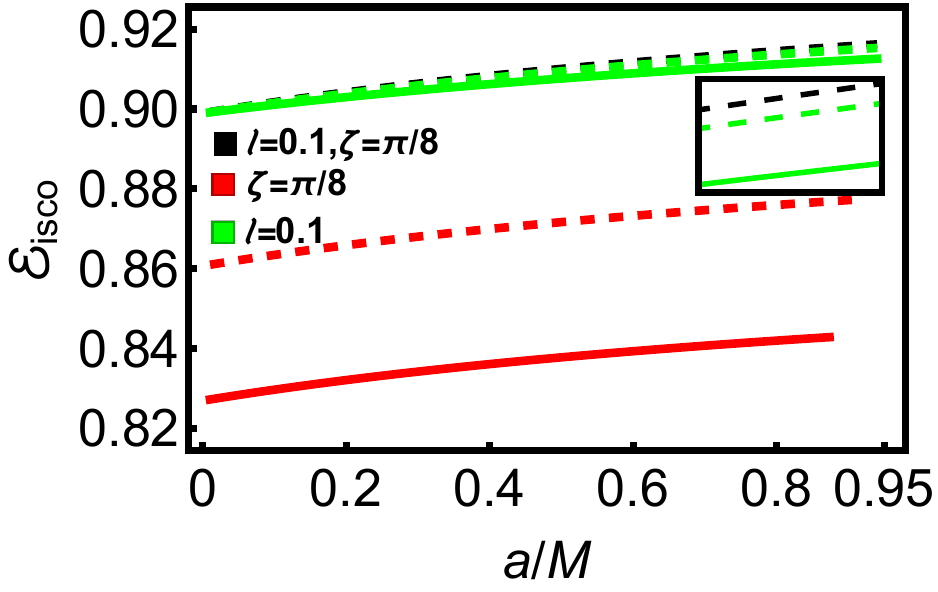}\label{neisco}}
\subfigure[]{\includegraphics[width=0.236\textwidth]{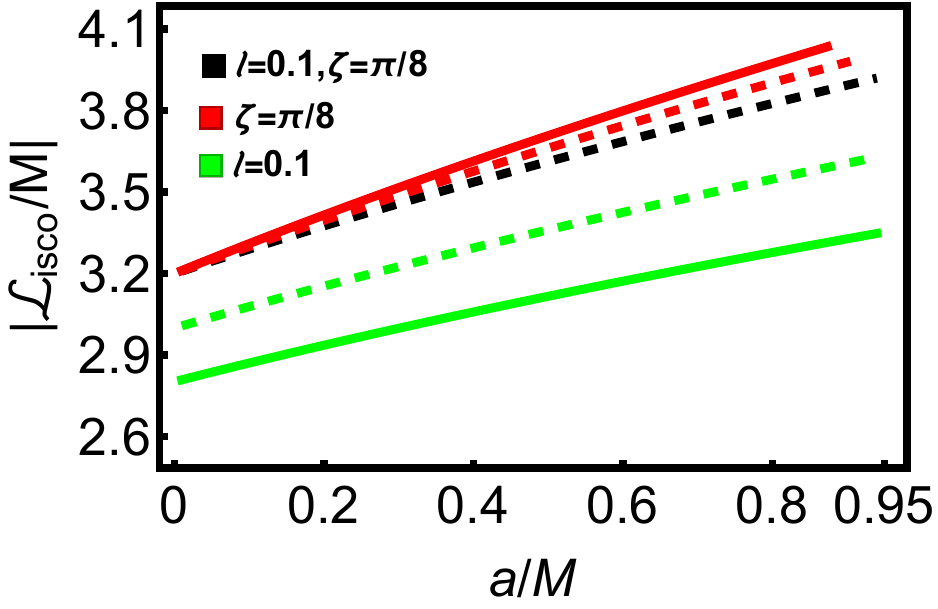}\label{nlisco}}
\subfigure[]{\includegraphics[width=0.225\textwidth]{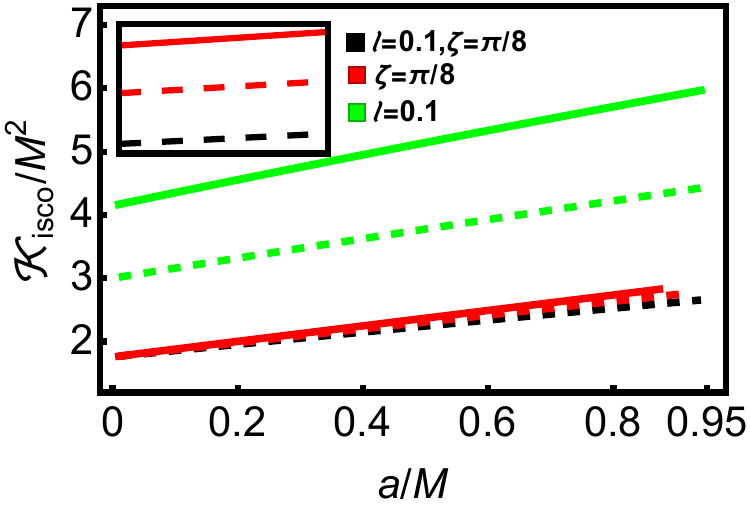}\label{ncisco}}
\caption{Characteristic quantities of the ISSO, including the radius $r_{ISSO}$, energy $\mathcal{E}_{ISSO}$, angular momentum $\mathcal{L}_{ISSO}$, and Carter constat $\mathcal{K}_{ISSO}$. Panels (a)–(d) depict the prograde orbits, while panels (e)–(h) present the retrograde orbits. The dashed and solid red curves correspond to $\ell=0.2~\text{and}~\ell=0.3$, the dashed and solid green curves correspond to $\zeta=\pi/6~\text{and}~\zeta=\pi/4$, respectively.}\label{figISSO}
\end{figure}

By considering various values of the LSB parameter $\ell$ and the tilt angle $\zeta$, we numerically solve the conditions \eqref{issocondition} to determine the radius, angular momentum, energy, and Carter constant of the innermost stable spherical orbits (ISSOs), as shown in Fig. \ref{figISSO}. For prograde ISSOs, all quantities, including the radius $r_{\text{ISSO}}$, angular momentum $\mathcal{L}_{\text{ISSO}}$, energy $\mathcal{E}_{\text{ISSO}}$, and Carter constant $\mathcal{K}_{\text{ISSO}}$, decrease with increasing black hole spin parameter $a$, whereas for retrograde ISSOs, the numerical values of these quantities increase as spin parameter $a$ grows. The LSB parameter $\ell$ suppresses all properties of prograde ISSOs, while the tilt angle $\zeta$ enhances $r_{\text{ISSO}}$, $\mathcal{E}_{\text{ISSO}}$, and $\mathcal{K}_{\text{ISSO}}$, but reduces $\mathcal{L}_{\text{ISSO}}$. Conversely, for retrograde ISSOs, the increasing LSB parameter $\ell$ enhances $r_{\text{ISSO}}$, $\mathcal{L}_{\text{ISSO}}$, and $\mathcal{K}_{\text{ISSO}}$, while reducing $\mathcal{E}_{\text{ISSO}}$; increasing $\zeta$ promotes $\mathcal{K}_{\text{ISSO}}$, but suppresses the radius $r_{\text{ISSO}}$, the angular momentum $\mathcal{L}_{\text{ISSO}}$, and the  energy $\mathcal{E}_{\text{ISSO}}$.

\section{Precession of spherical orbits}\label{secPOSO}

For the equatorial circular orbits, the orbital angular momentum is aligned with the black hole spin, constraining the motion to the equatorial plane. In contrast, when the orbit is inclined, the orbital angular momentum forms a nonzero angle with the spin axis, leading to different orbital periods in the $\theta$ and $\phi$ directions, and gives rise to Lense–Thirring precession. In this section, we investigate the precession of spherical orbits by analyzing the effects of the parameters ($a,\ell,\zeta$) on the evolution of motion in the $\theta$ and $\phi$ directions.

To characterize the precession of spherical orbits, we examine the evolution of the angular coordinates $\theta$ and $\phi$ with respect to the coordinate time $t$. From Eqs. \eqref{eqttl} and \eqref{equtheta}, we obtain 
\begin{align}
\frac{d\theta}{dt}=&\frac{\pm\sqrt{\Theta(\theta)}}{a\left(\mathcal{L}-a\mathcal{E}\sin^{2}\theta\right)+\frac{\left(r^{2}+a^{2}\right)\left(r^{2}\mathcal{E}+a^{2}\mathcal{E}-a\mathcal{L}\right)}{\Delta}}\label{eqdthetadt},\\
\frac{d\phi}{dt}=&\frac{\left(\mathcal{L}\csc^{2}\theta-a\mathcal{E}\right)+\frac{a\left(r^{2}\mathcal{E}+a^{2}\mathcal{E}-a\mathcal{L}\right) }{\Delta}}{a\left(\mathcal{L}-a\mathcal{E}\sin^{2}\theta\right)+\frac{\left(r^{2}+a^{2}\right)\left(r^{2}\mathcal{E}+a^{2}\mathcal{E}-a\mathcal{L}\right)}{\Delta}}\label{eqdphidt},
\end{align}
where $\mathcal{E}$ and $\mathcal{L}$ are solved in Eqs. \eqref{sporbit}.

\begin{figure}[!htb]
\subfigure[]{\includegraphics[width=0.236\textwidth]{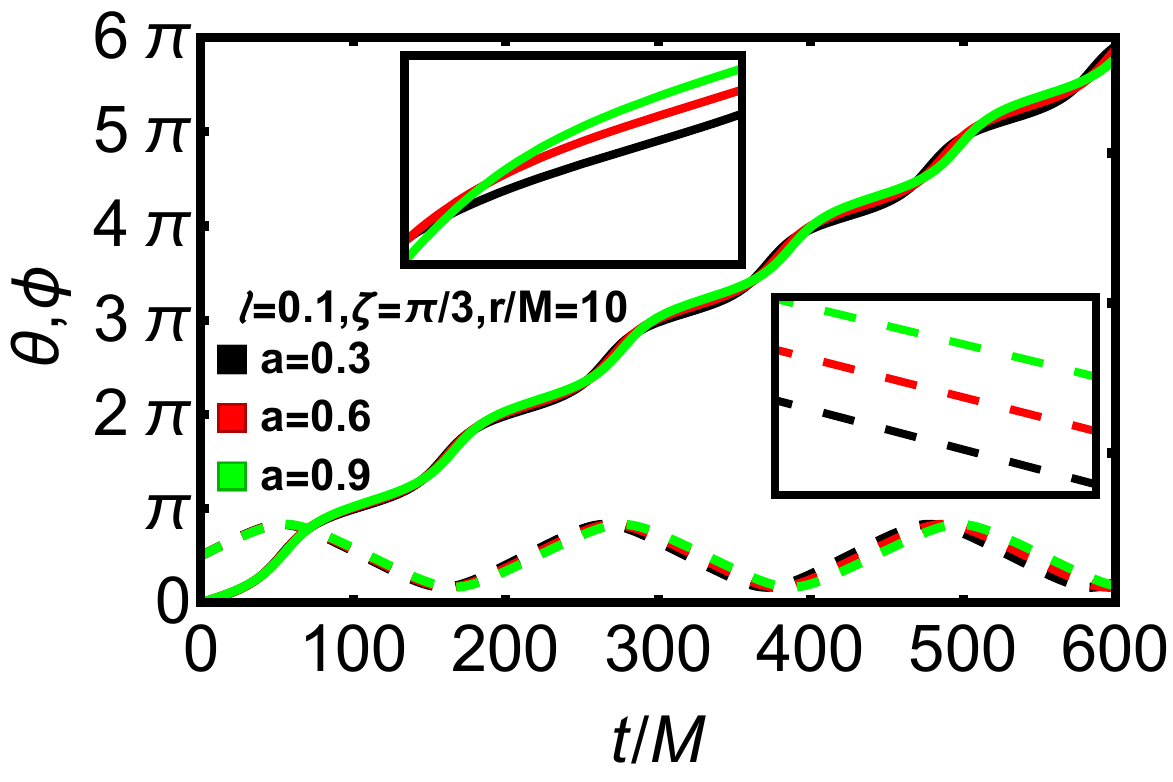}\label{paevolution}}
\subfigure[]{\includegraphics[width=0.236\textwidth]{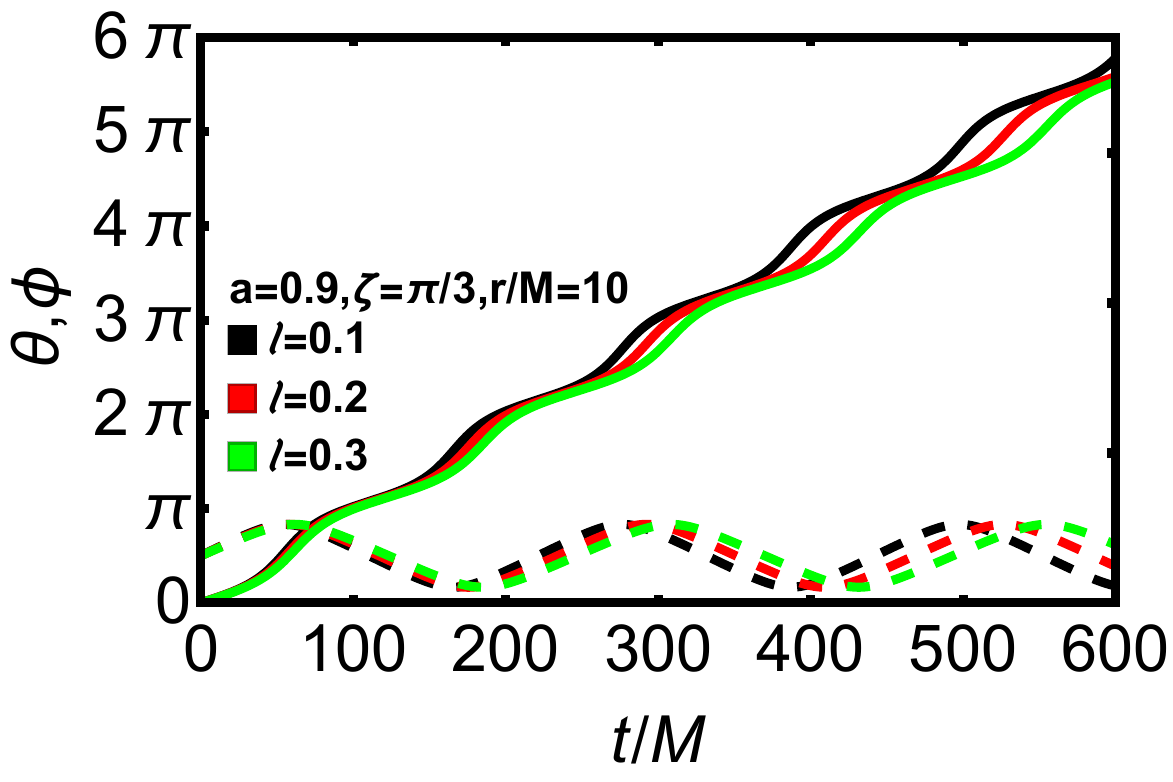}\label{plevolution}}
\subfigure[]{\includegraphics[width=0.236\textwidth]{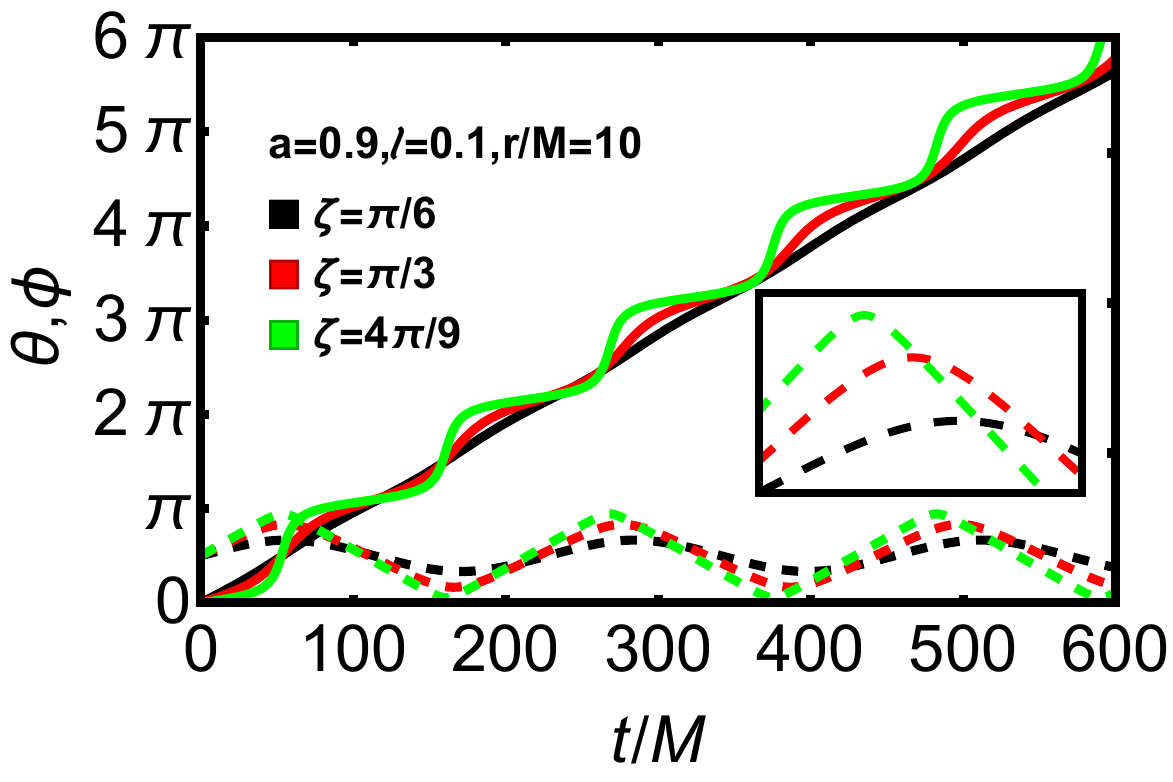}\label{pzevolution}}
\subfigure[]{\includegraphics[width=0.236\textwidth]{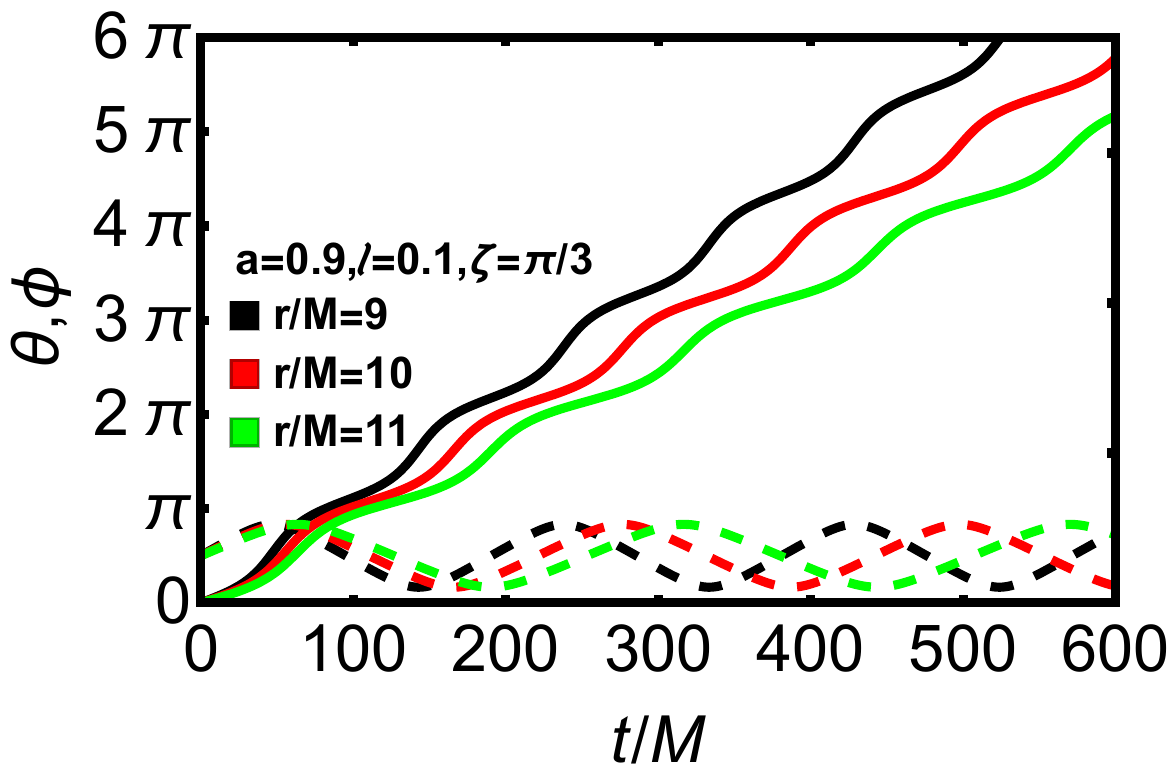}\label{prevolution}}\\
\subfigure[]{\includegraphics[width=0.236\textwidth]{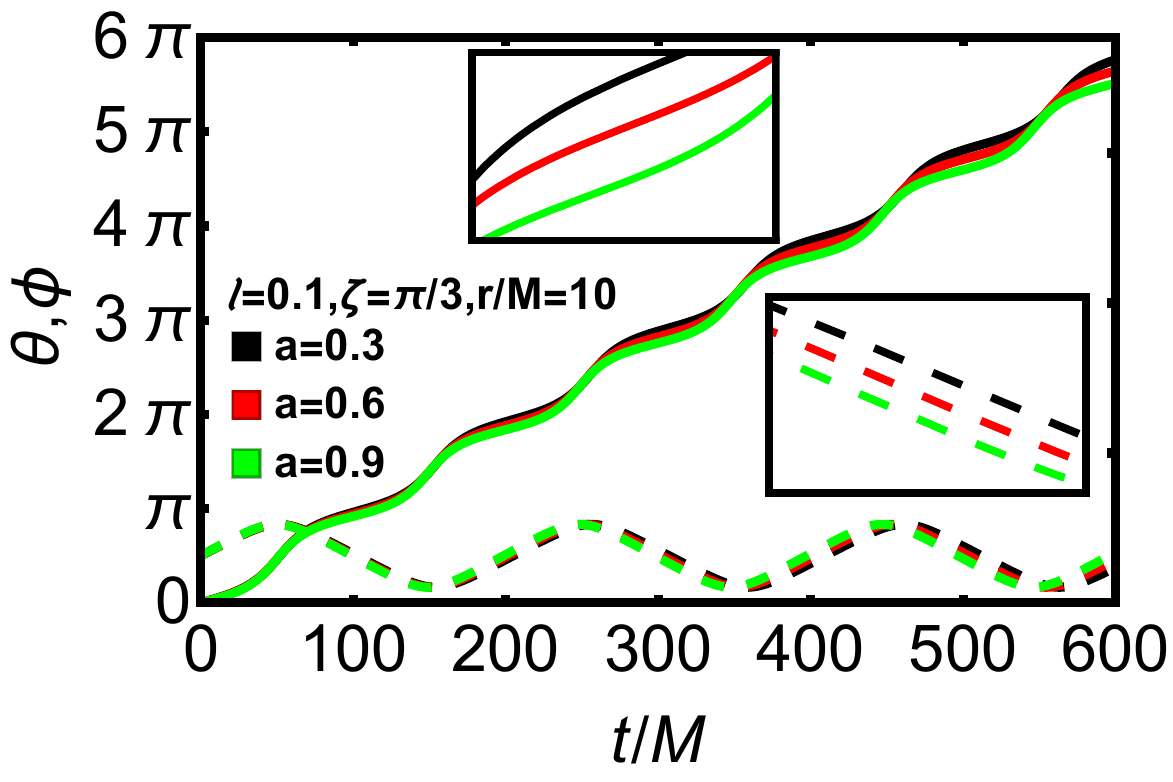}\label{naevolution}}
\subfigure[]{\includegraphics[width=0.236\textwidth]{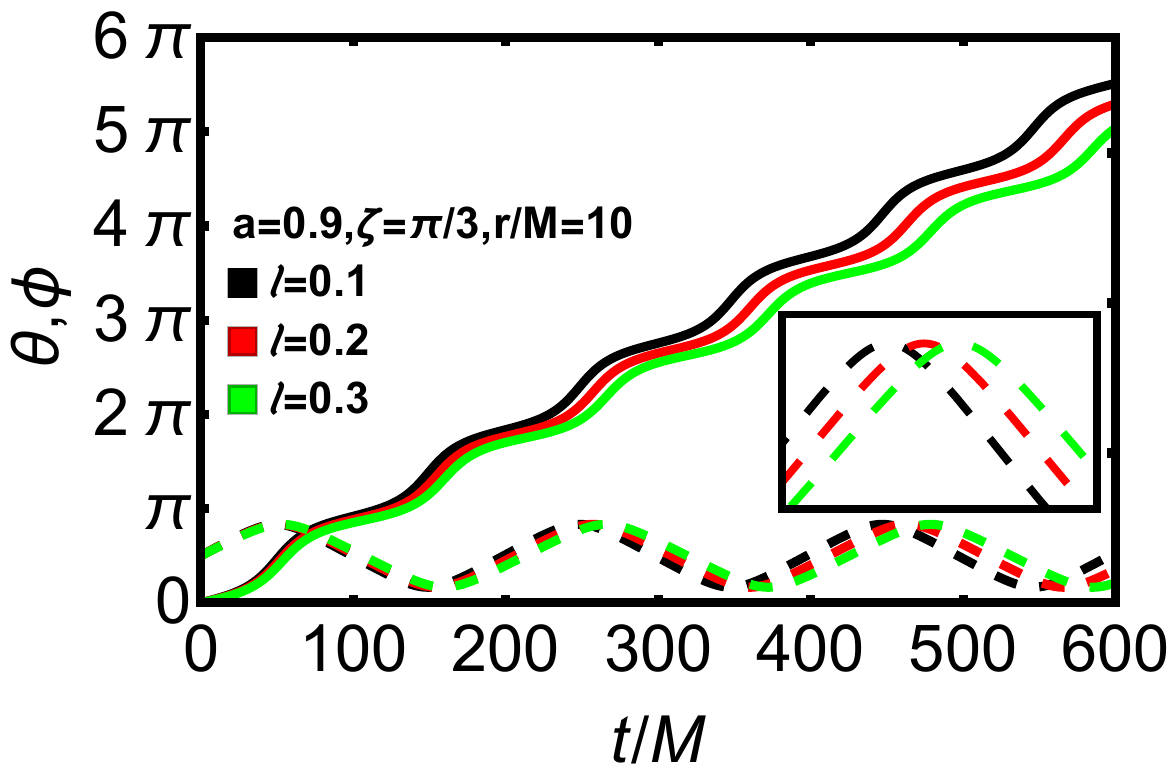}\label{nlevolution}}
\subfigure[]{\includegraphics[width=0.236\textwidth]{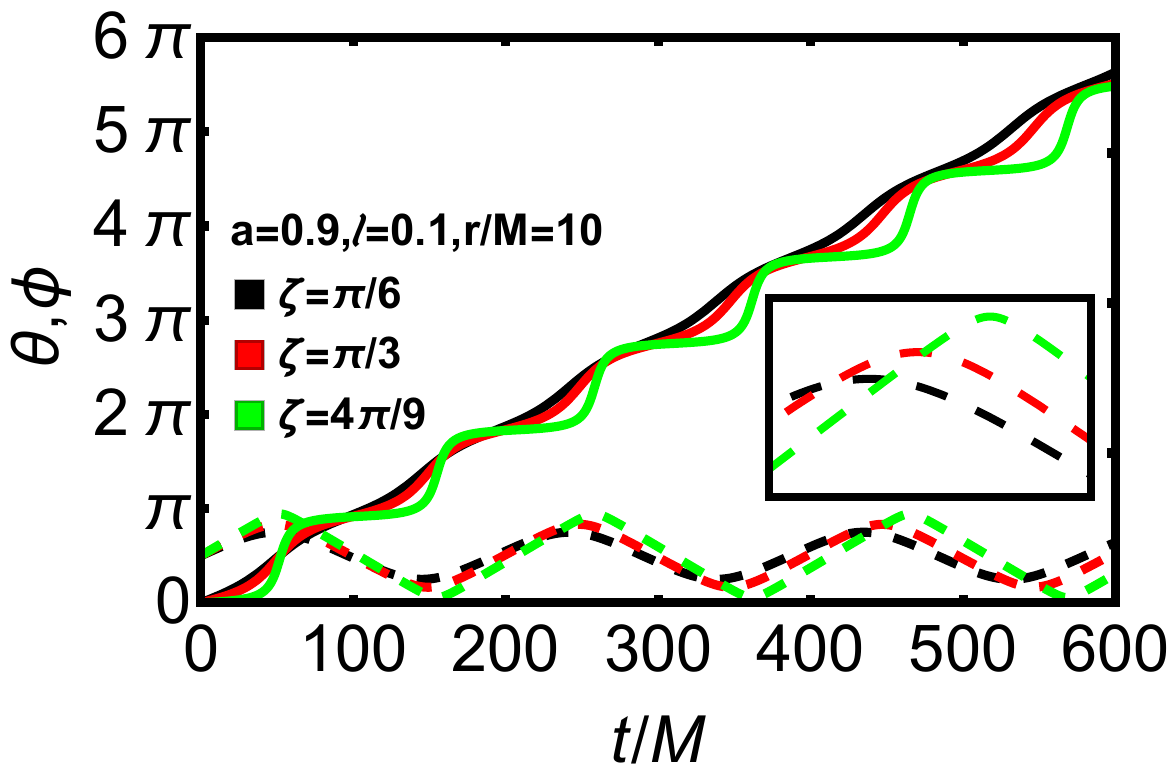}\label{nzevolution}}
\subfigure[]{\includegraphics[width=0.236\textwidth]{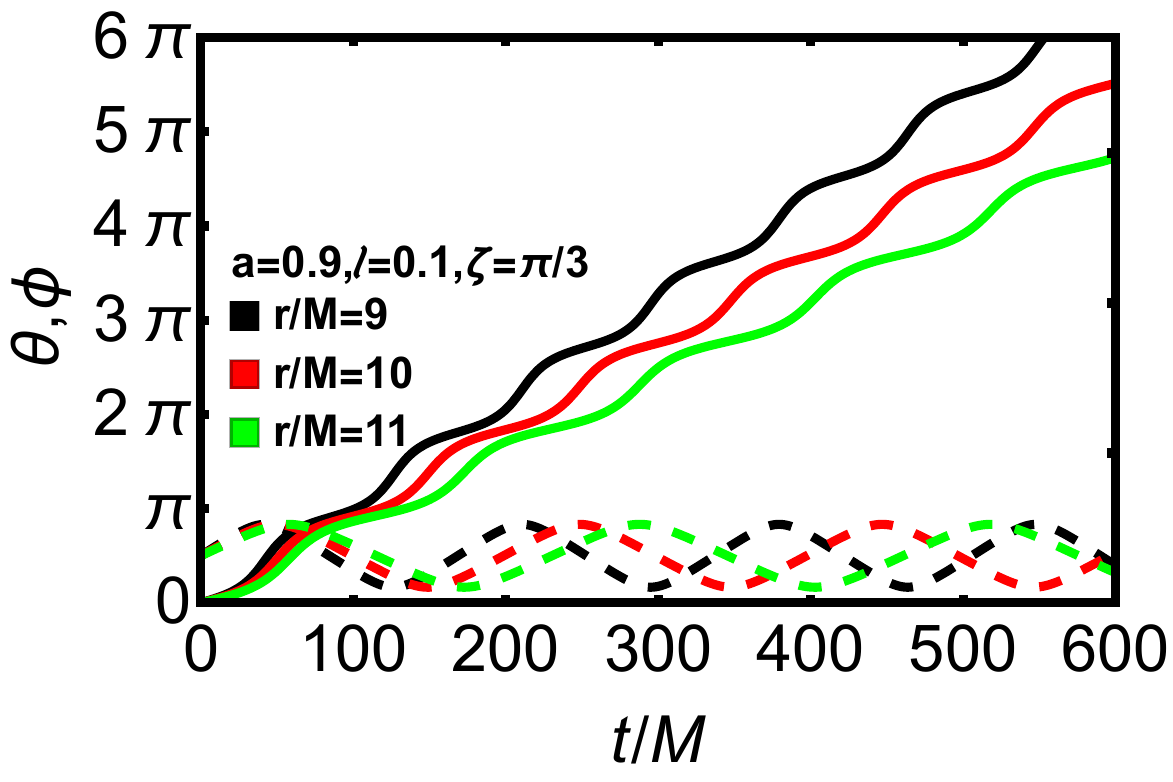}\label{nrevolution}}
\caption{Evolution of the angular coordinates $\theta$ and $\phi$ as functions of the coordinate time $t$. Panels (a)–(d) correspond to prograde spherical orbits, while panels (e)–(h) illustrate retrograde spherical orbits.}\label{evolution}
\end{figure}

With the initial conditions $\theta(0)=\frac{\pi}{2}$ and $\phi(0)=0$, the evolution of the angular coordinates $\theta$ and $\phi$ can be obtained by numerically integrating the corresponding equations of motion. The resulting trajectories are shown in Fig. \ref{evolution}, which illustrates the time evolution of both angular variables. For a complementary three-dimensional visualization of the motion of a massive particle on a spherical orbit, see Fig. 3 of Ref. \cite{AlZahrani:2023xix}. 

From Fig. \ref{evolution}, the motion in the $\theta$ direction exhibits periodic oscillations in coordinate time $t$, confined within the range $(\pi/2-\zeta,\pi/2+\zeta)$. In contrast, the evolution of $\phi$ increases monotonically without bound and is approximately linear with small deviations, indicating that the particle maintains an effectively constant azimuthal velocity. As shown in Figs. \ref{pzevolution} and \ref{nzevolution}, both the oscillation amplitude in $\theta$ and the deviation from linearity in $\phi$ increase as the tilt angle $\zeta$ grows. A more detailed discussion of the influence of larger $\zeta$ on the behavior of $\theta(t)$ and $\phi(t)$ is provided in Sec. III of Ref. \cite{Wei:2024cti}.

\begin{table*}
\centering
\addtolength\tabcolsep{5pt}
\begin{tabular}{c|cc|cc|cc|cc}\hline \hline 
\multirow{2}*{prograde} &\multicolumn{2}{c}{Fig.(a)} &\multicolumn{2}{c}{Fig.(b)} &\multicolumn{2}{c}{Fig.(c)} &\multicolumn{2}{c}{Fig.(d)}\\\cline{2-9}
~~&$T_\theta$ & $\phi(T_\theta)/\pi$& $T_\theta$ & $\phi(T_\theta)/\pi$&$T_\theta$ & $\phi(T_\theta)/\pi$&$T_\theta$ & $\phi(T_\theta)/\pi$\\\hline 
$\text{black}$ & $212.3$ & $2.053$ ~~ & $221.1$ & $2.154$
               & $227.4$ & $2.147$ ~~ & $190.4$ & $2.175$\\\hline
$\text{red}$   & $216.6$ & $2.105$ ~~ & $233.9$ & $2.201$ 
               & $221.1$ & $2.154$ ~~ & $221.1$ & $2.154$\\\hline
$\text{green}$ & $221.1$ & $2.154$ ~~ & $246.6$ & $2.246$
               & $214.4$ & $2.160$ ~~ & $253.3$ & $2.138$\\
\hline \hline 
\multirow{2}*{retrograde}&\multicolumn{2}{c}{Fig.(e)} &\multicolumn{2}{c}{Fig.(f)} &\multicolumn{2}{c}{Fig.(g)} &\multicolumn{2}{c}{Fig.(h)}\\\cline{2-9}
~ & $T_\theta$ & $\phi(T_\theta)/\pi$& $T_\theta$ & $\phi(T_\theta)/\pi$ & $T_\theta$ & $\phi(T_\theta)/\pi$&  $T_\theta$ & $\phi(T_\theta)/\pi$\\\hline 
$\text{black}$ & $204.8$ & $1.945$ ~~ & $198.6$ & $1.832$
               & $189.1$ & $1.829$ ~~ & $168.5$ & $1.810$\\\hline
$\text{red}$   & $201.5$ & $1.889$ ~~ & $205.8$ & $1.789$ 
               & $198.6$ & $1.832$ ~~ & $198.6$ & $1.832$\\\hline
$\text{green}$ & $198.6$ & $1.832$ ~~ & $212.7$ & $1.750$
               & $206.6$ & $1.835$ ~~ & $230.2$ & $1.849$\\\hline
\end{tabular}
\caption{The period $T_\theta$ of the motion in $\theta$, and the angular rotation of $\phi(T_\theta)/\pi$ within the cycle of $T_\theta$.}\label{table1}
\end{table*}

Furthermore, Tab. \ref{table1} lists the period $T_\theta$ of the motion in $\theta$, along with the corresponding angular displacement $\phi(T_\theta)/\pi$ accumulated in the $\phi$ direction over one $\theta$-cycle, for the configurations shown in Fig. \ref{evolution}. The result shows that, for prograde orbits, increases in both the spin parameter $a$ and the LSB parameter $\ell$ lead to larger values of $T_\theta$ and the angle swept $\phi(T_\theta)/\pi$. By contrast, increasing the tilt angle $\zeta$ reduces the period $T_\theta$, while enhancing the accumulated angle $\phi(T_\theta)/\pi$. A larger orbital radius $r/M$ increases the oscillation period $T_\theta$, but suppresses $\phi(T_\theta)/\pi$. On the other hand, for the retrograde orbits, an increase in the spin parameter $a$ reduces $T_\theta$, whereas increases in ($\ell,\zeta,r/M$)  all act to lengthen the period. Additionally, increase in both $a$ and $\ell$ suppresses the rotational angle $\phi(T_\theta)/\pi$, whereas the increase in $\zeta$ and $r/M$ have pronounced effect on $\phi(T_\theta)/\pi$.

\begin{figure}[!h]
\subfigure[]{\includegraphics[width=0.236\textwidth]{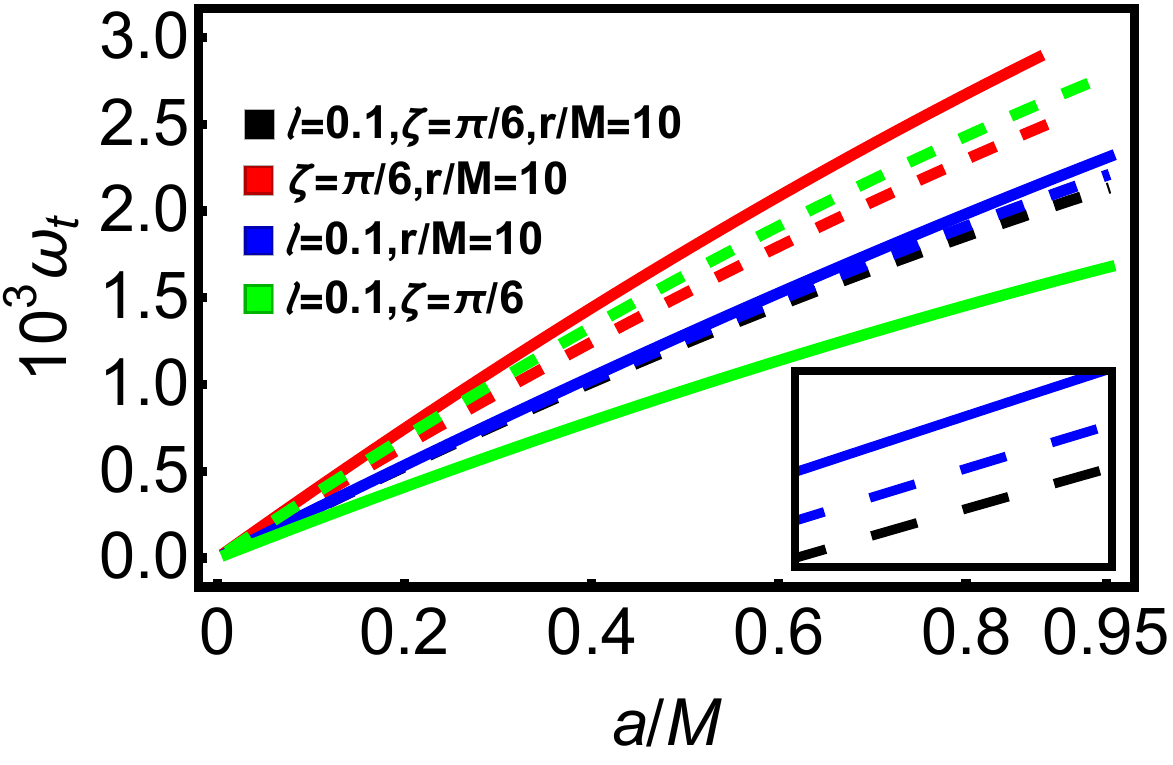}\label{pvelocitya}}
\subfigure[]{\includegraphics[width=0.236\textwidth]{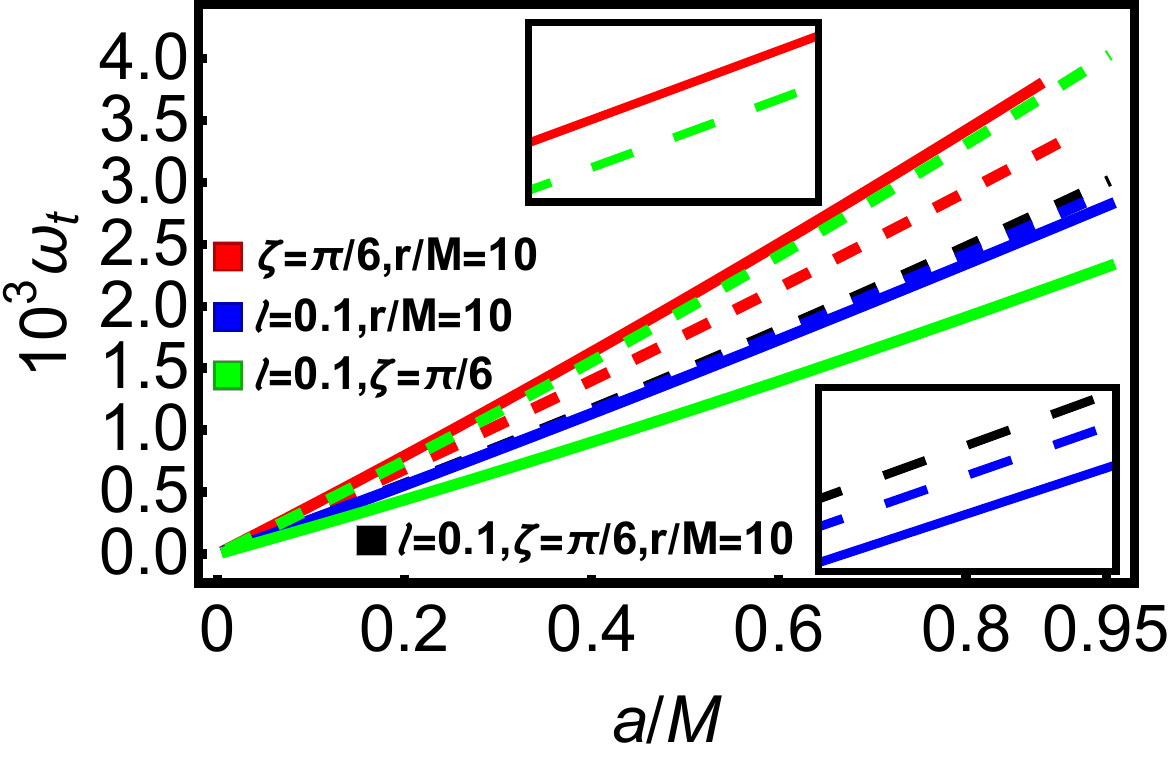}\label{nvelocitya}}
\caption{Precession angular velocity $\omega_{t}$ as a function of the spin parameter $a$ for various values of ($\ell,\zeta,r/M$). 
The left panel corresponds to prograde spherical orbits, while the right panel shows retrograde ones. The dashed and solid red curves represent $\ell=0.2~\text{and}~\ell=0.3$, respectively; the dashed and solid blue curves denote $\zeta=\pi/4~\text{and}~\zeta=\pi/3$; and the dashed and solid green curves correspond to $r/M=9~\text{and}~r/M=11$, respectively.}\label{figomegat}
\end{figure}

It is noteworthy that after one full cycle of duration $T_\theta$, the $\phi$-motion does not return to $2\pi$, and this deviation characterizes the precession angular velocity $\omega_t$ of the spherical orbit. In Ref. \cite{AlZahrani:2023xix}, $\omega_t$ was defined through the transformation $\phi(t)\rightarrow\phi(t)-\omega_t t$, allowing the orbit to appear stationary and perfectly circular, as illustrated in their Fig. 4. In this work, we compute the precession angular velocity $\omega_{t}$ following the definition in Ref. \cite{Wei:2024cti}, given by the expression with 
\begin{align}
\label{eqomega}\omega_{t}=\frac{|\phi(T_\theta)-2\pi|}{T_\theta},
\end{align}
and the numerical results corresponding to different parameter values of ($a, \ell,\zeta, r/M$) show in Fig. \ref{figomegat}.

The results indicate that increases in both the spin parameter $a$ and LSB parameter $\ell$ enhance the precession angular velocity $\omega_{t}$ for both prograde and retrograde spherical orbits. Along prograde orbits, the precession angular velocity $\omega_{t}$ grows with the tilt angular $\zeta$, whereas for retrograde orbits it decreases as $\zeta$ increases. A larger orbital radius $r/M$ suppresses the value of $\omega_{t}$. Moreover, it is worth emphasizing that, compared with the influence of other parameters, the effect of the tilt angle $\zeta$ on $\omega_{t}$ is relatively weak, consistent with the findings reported in Ref. \cite{AlZahrani:2023xix}.

\section{Precession period and Constrains of M87*}\label{secConstrain}

The shadow image of the supermassive black hole M87* as crescent shaped was photographed by the EHT collaboration with the mass $M=6.5\times10^9M_\odot$, and determined that the angular radius of the shadow  with $\theta_{sh}=42\pm3\mu as$ \cite{EventHorizonTelescope:2019dse,EventHorizonTelescope:2019uob,EventHorizonTelescope:2019jan}. Then, based on an analysis of 22 years of radio observational data from the M87*, Cui et al. discovered a periodic variation of approximately $T=11.24\pm0.47$ years with the half-opening angle of the precession cone $1.25^\circ \pm 0.18^\circ$, attributing this phenomenon to Lense-Thirring precession induced by the misalignment between the accretion disk and the black hole spin \cite{Cui:2023uyb}. This precessional motion provides a valuable means of constraining black hole parameters through the characteristic frequencies of spherical orbital motion. Furthermore, general relativistic magnetohydrodynamic (GRMHD) simulations suggest that in misaligned disk–jet systems, a slight tilt between the rotational axis of the accretion disk and the spin axis of the black hole can lead to substantial Lense–Thirring precession of a significant portion of the disk \cite{Fragile:2007dk,Liska:2017alm,White:2019udt,Chatterjee:2020eqc,Ressler:2023ptc}, with the jet precessing coherently alongside the disk \cite{Liska:2017alm,McKinney:2012wd}.

In realistic accretion systems, the precession of jets or accretion disks is influenced by a combination of factors, including disk viscosity, magnetic fields, and gravitational perturbations from external bodies. In this work, we adopt a simplified model, assuming that the precession frequency under idealized conditions is decoupled from the effects of external accretion flows and surrounding celestial objects. We focus exclusively on the fundamental dynamics of the accretion disk, employing spherical orbits to model particle motion within the disk. During this initial modeling stage, treating the precession frequency as dependent solely on the black hole spin and the orbital structure allows for a significant simplification, providing a tractable and reasonably accurate framework for estimating black hole parameters prior to more detailed simulations. The precession period, after restoring physical units, can be expressed as follows:
\begin{align}\label{eqperiod}
T=\frac{2\pi}{\omega_{t}}\frac{GM}{c^{3}}\approx 6.372\times10^{-3}\times\frac{1}{\omega_{t}}(\text{year}),
\end{align}
where we have set $M=6.5\times10^9 M_{\odot}$ with the mass of M87* black hole. 

The Eqs. \eqref{eqdthetadt} and \eqref{eqdphidt} indicate that the precession angular velocity $\omega_{t}$ depends on the parameters as $\omega_{t}=\omega_{t}(r, a, \ell, \zeta)$. Furthermore, combining the results in Fig. \ref{figomegat} with Ref. \cite{AlZahrani:2023xix}, it is evident that the influence of the tilt angle $\zeta$ on $\omega_t$ can be negligible. To simplify the parameter constraints $(a,\ell,r/M)$, we therefore fix the tilt angle at $\zeta_{p}=1.25^\circ$. It is also important to note that current observations do not distinguish between prograde and retrograde orbits; accordingly, we account for potential differences between the two orbit types in our analysis. 

On the other hand, EHT observations of the shadow and jet suggest that M87* likely possesses a geometrically extended accretion disk, indicating that modeling the disk using the ISSO alone may be inadequate. Studies of inclined accretion disks have further shown that the ISSO does not characterize the entire disk \cite{Ostriker1989MNRAS}, as particles within the ISSO rapidly plunge into the black hole. Previous investigations \cite{Fragile:2000mf,Lodato_2010} have demonstrated that the disk inclination decreases with decreasing radius and the disk realigns with the equatorial plane at a characteristic warp radius, . Accordingly, it is reasonable to assume that jets are launched from the warp radius, which lies outside the ISSO. Motivated by this, we focus on examining the precession period of spherical orbits at various radii within the accretion disk $r>r_{ISSO}$ and explore its potential observational implications.
 
\begin{figure}[!htb]
\subfigure[ prograde orbit]{\includegraphics[width=0.236\textwidth]{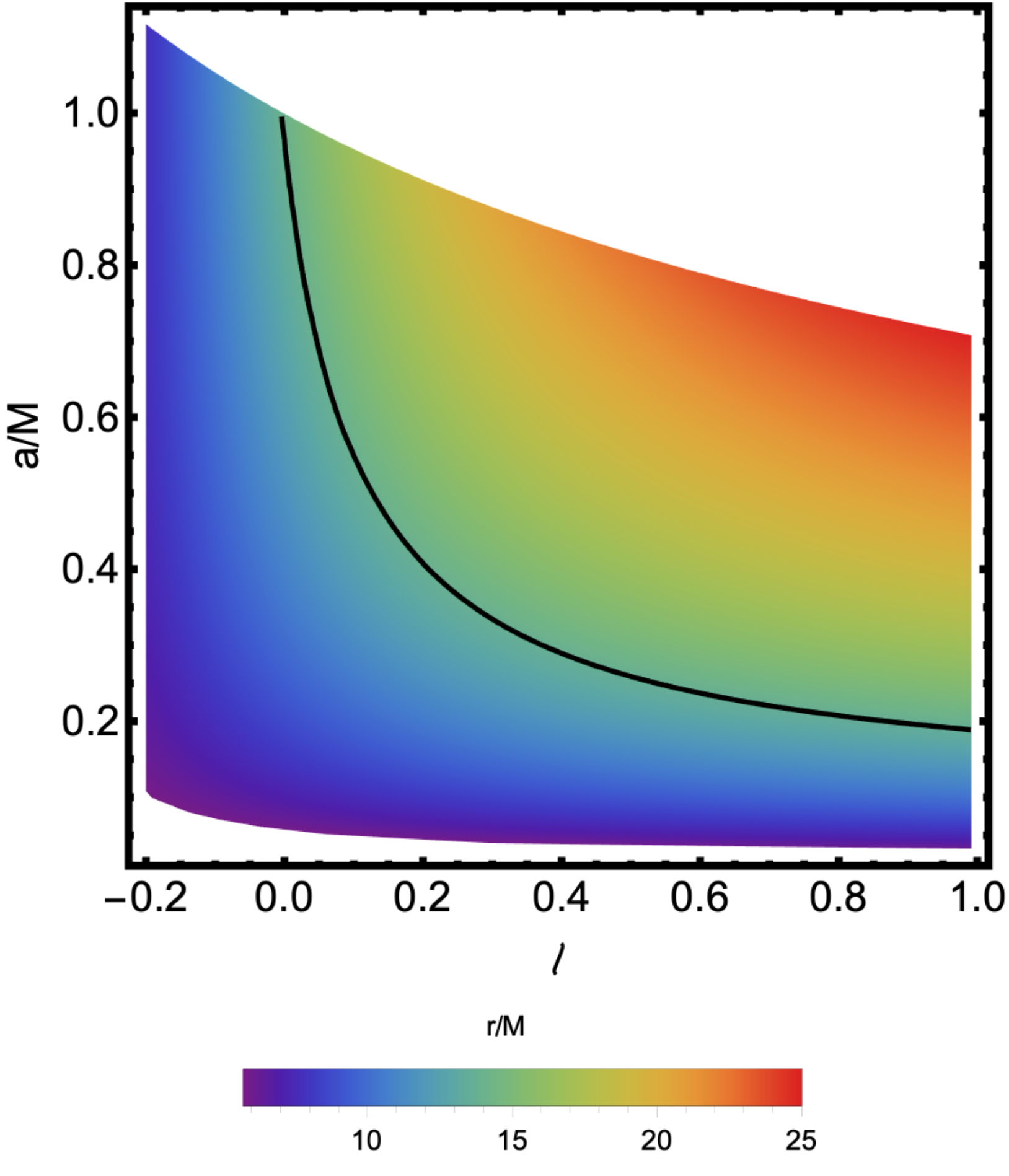}\label{ptlimit}}
\subfigure[$\Delta P(r)$ ]{\includegraphics[width=0.236\textwidth]{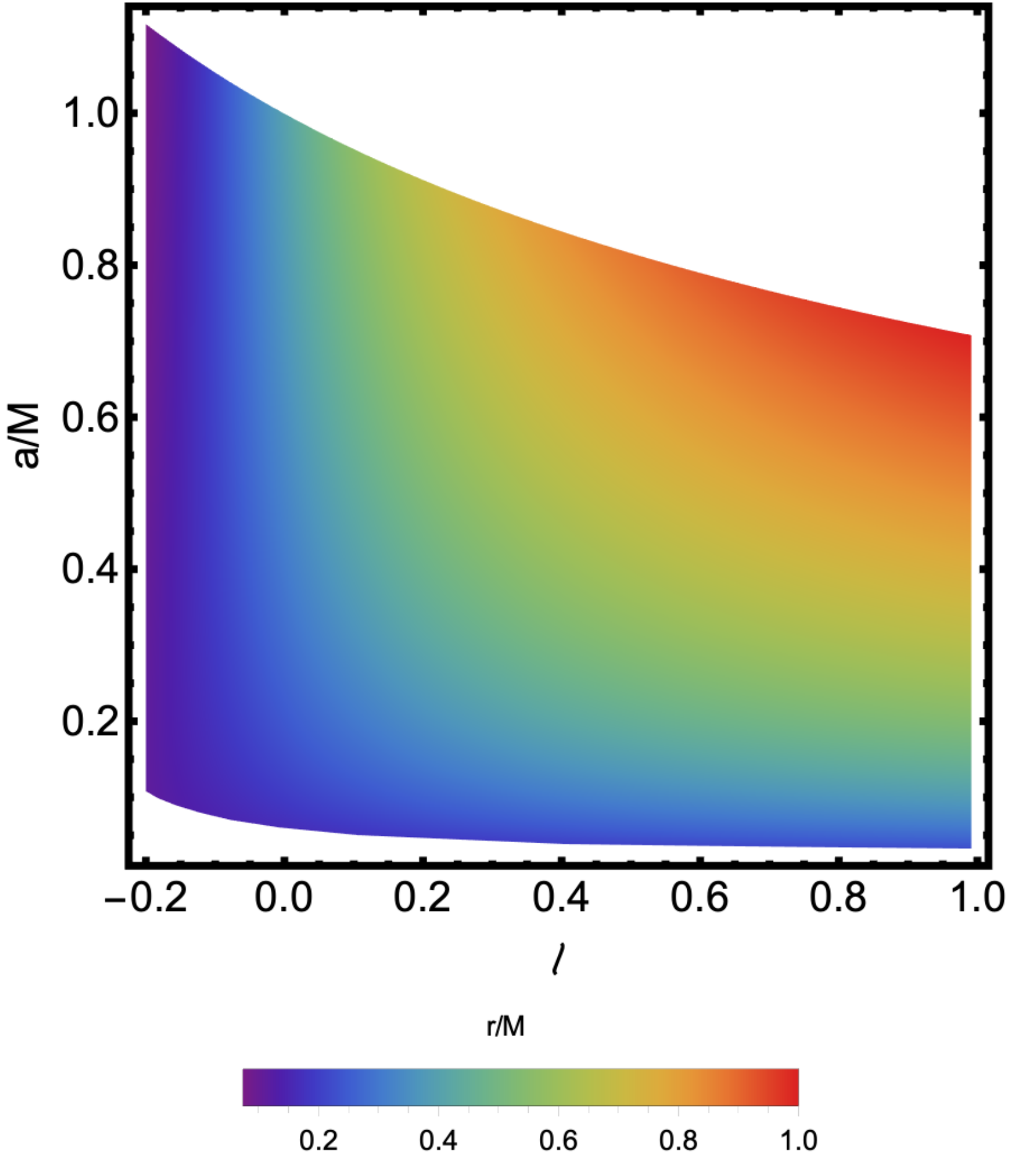}\label{dplimitrl}}
\subfigure[ retrograde orbit]{\includegraphics[width=0.236\textwidth]{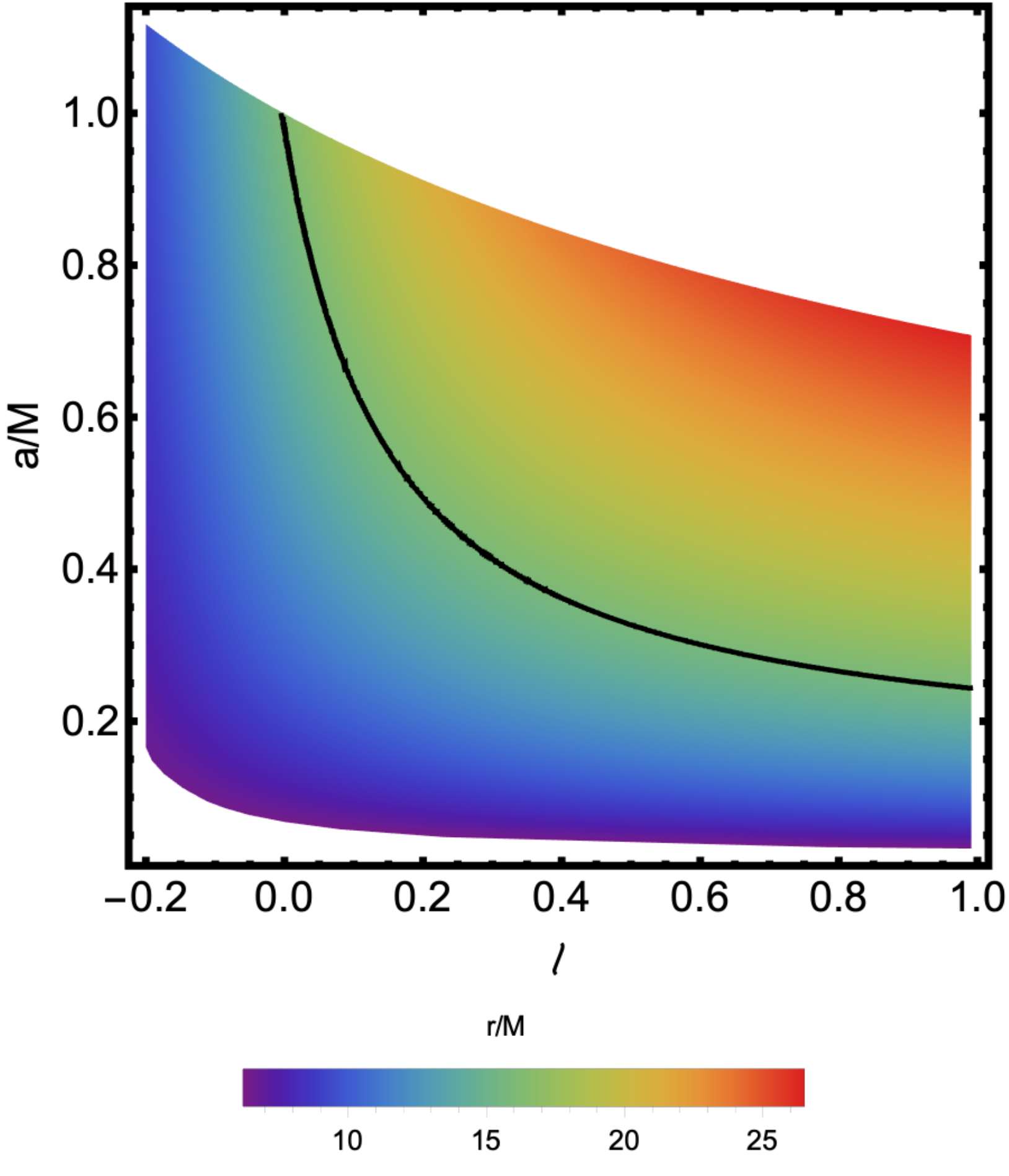}\label{ntlimit}}
\subfigure[$\Delta R(r)$]{\includegraphics[width=0.236\textwidth]{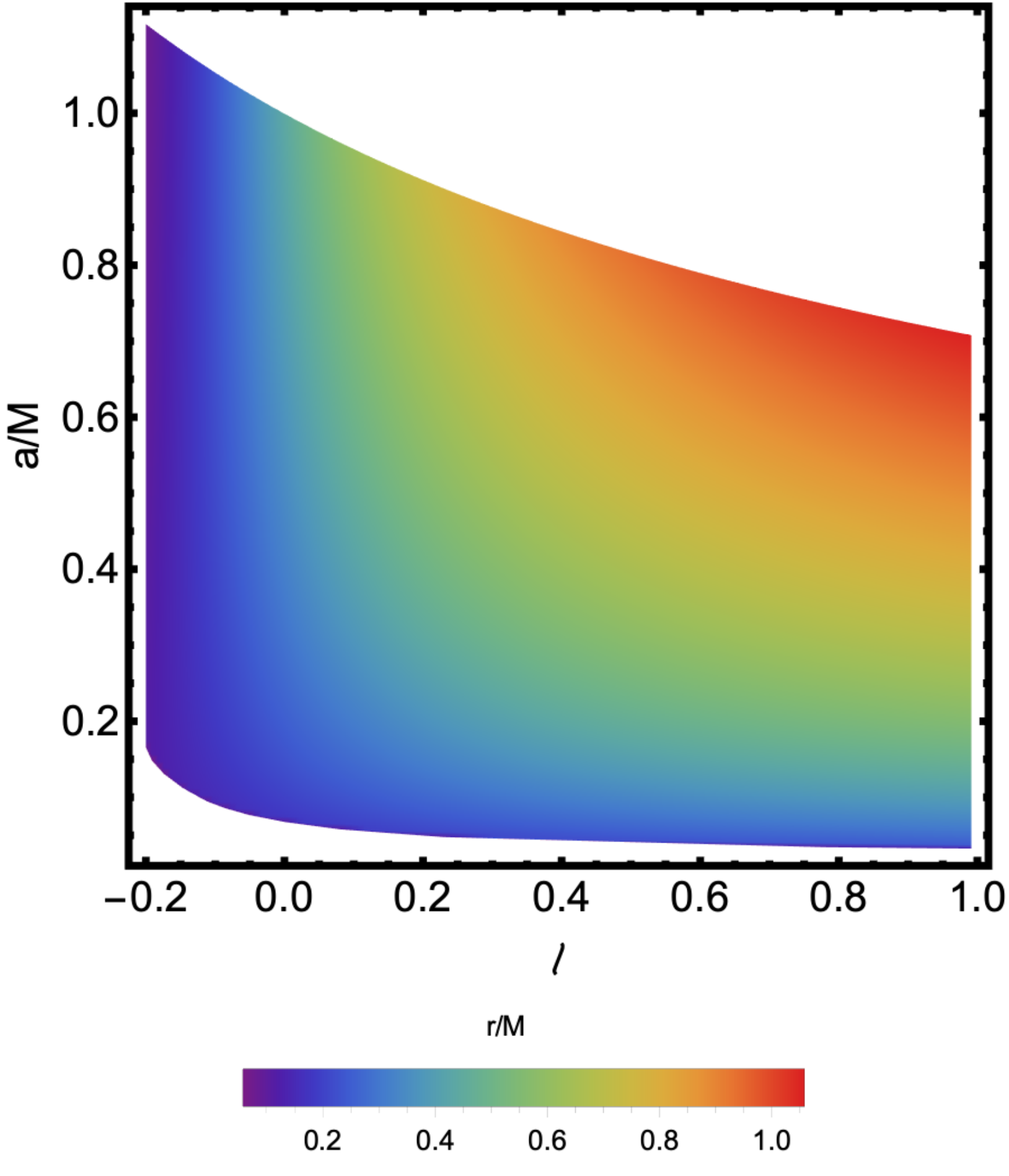}\label{dnlimitrl}}
\subfigure[$\Delta RP(r)$]{\includegraphics[width=0.236\textwidth]{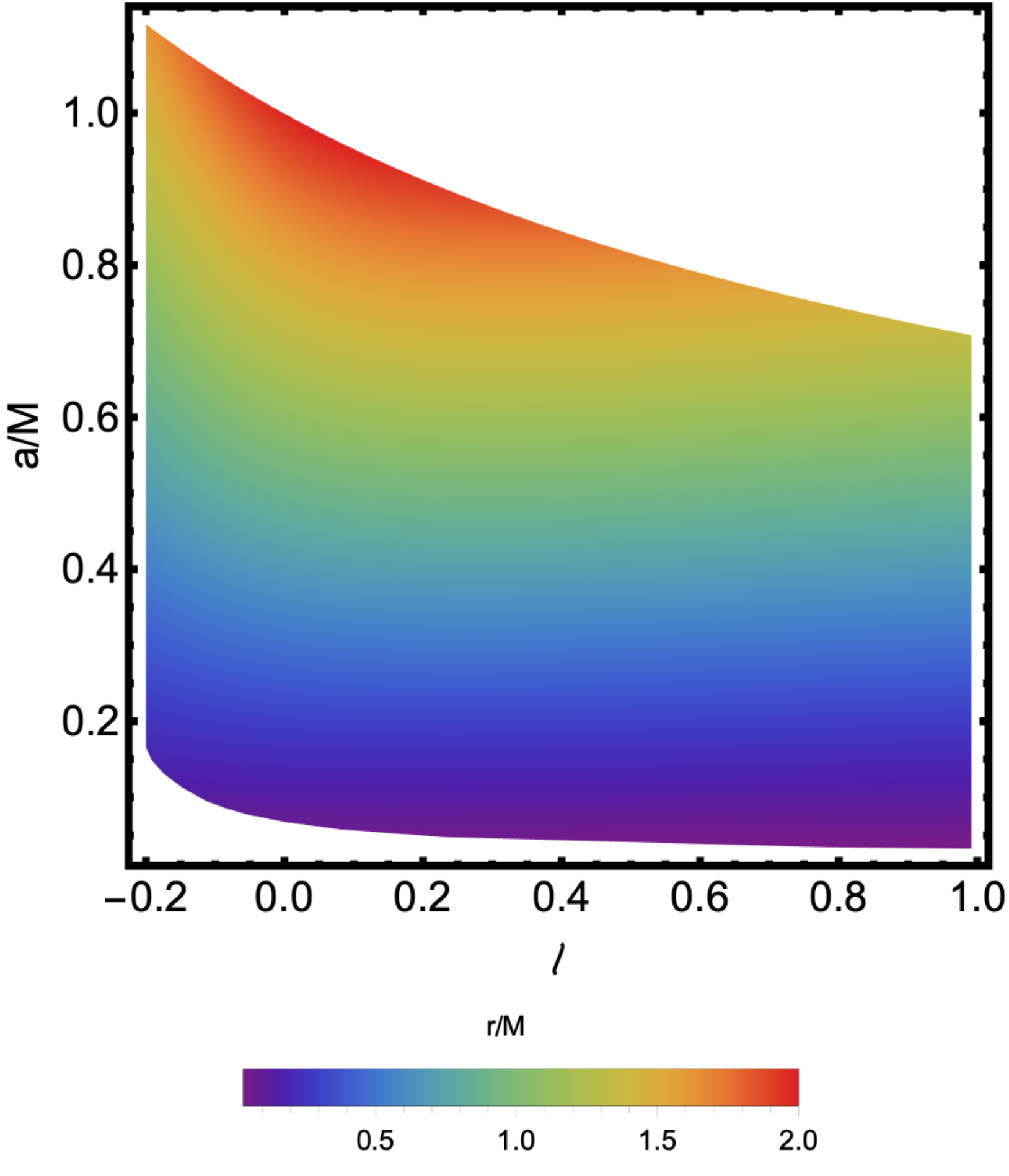}\label{dnplimit}}
\caption{Warp radius $r/M$ of M87$^{*}$ constrainted with $T=11.24\pm0.47$ year. $\Delta P(r)$ and $\Delta R(r)$ denote the discrepancy of the warp radius $r/M$ corresponding to the precession period limits $T = 10.77$ and $T = 11.71$ years, for the prograde orbit and the retrograde orbit, respectively. $\Delta RP(r)$ represents the difference in warp radius $r/M$  between the prograde orbit and the retrograde orbit with $T=11.24$ year.}\label{limtfit}
\end{figure}

Based on the result in Ref. \cite{Cui:2023uyb} that the precession period of the M87* jet as $11.24 \pm 0.47$ year, for a given set of black hole parameters $(a, \ell)$, we can determine the corresponding orbital radius $r/M$, and the result are exhibited with the $(a, \ell)$ planet density plots in Fig. \ref{limtfit}. The results show that with the precession period $T=11.24$ year, the warp radius $r/M$ spans $(5.73,25.15)$ for prograde orbits and $(6.16,26.46)$ for retrograde orbits. Increases in either the spin parameter $a$ or LSB parameter $\ell$ facilitate larger warp radii $r/M$, indicating that higher values of $a$ or $\ell$ correspond to more extended jet trajectories. The black lines in Figs. \ref{ptlimit} and \ref{ntlimit} denote the maximum orbital radii, ($r/M=14.12$) for prograde and ($r/M=16.1$) for retrograde orbits, respectively, corresponding to the Kerr black hole case as reported in Ref. \cite{Wei:2024cti}. This implies that when the warp radius $r/M$ exceeds 16, there may exist a non-vacuum bumblebee vector field in the vicinity of the black hole. Figures \ref{dplimitrl} and \ref{dnlimitrl} shows the variation of the warp radius $r/M$ between the constraints $T=10.77$ and $T=11.71$ year, for the prograde orbit and the retrograde orbit both fall within the domain of $(0.08,1.03)$.  Figure \ref{dnplimit} show the difference in warp radius $r/M$ between prograde and retrograde orbits under the constraint $T=11.24$, which ranges from $0.03$ to $1.99$. These results reveal that the spin parameter $a$ exerts a more significant influence on the warp radius $r/M$ than the LSB parameter $\ell$. In other words, only a larger spin parameter $a$ can facilitate the differentiation of jet motion on prograde or retrograde orbits. 

\begin{figure}[!htb]
\subfigure[prograde orbit]{\includegraphics[width=0.23\textwidth]{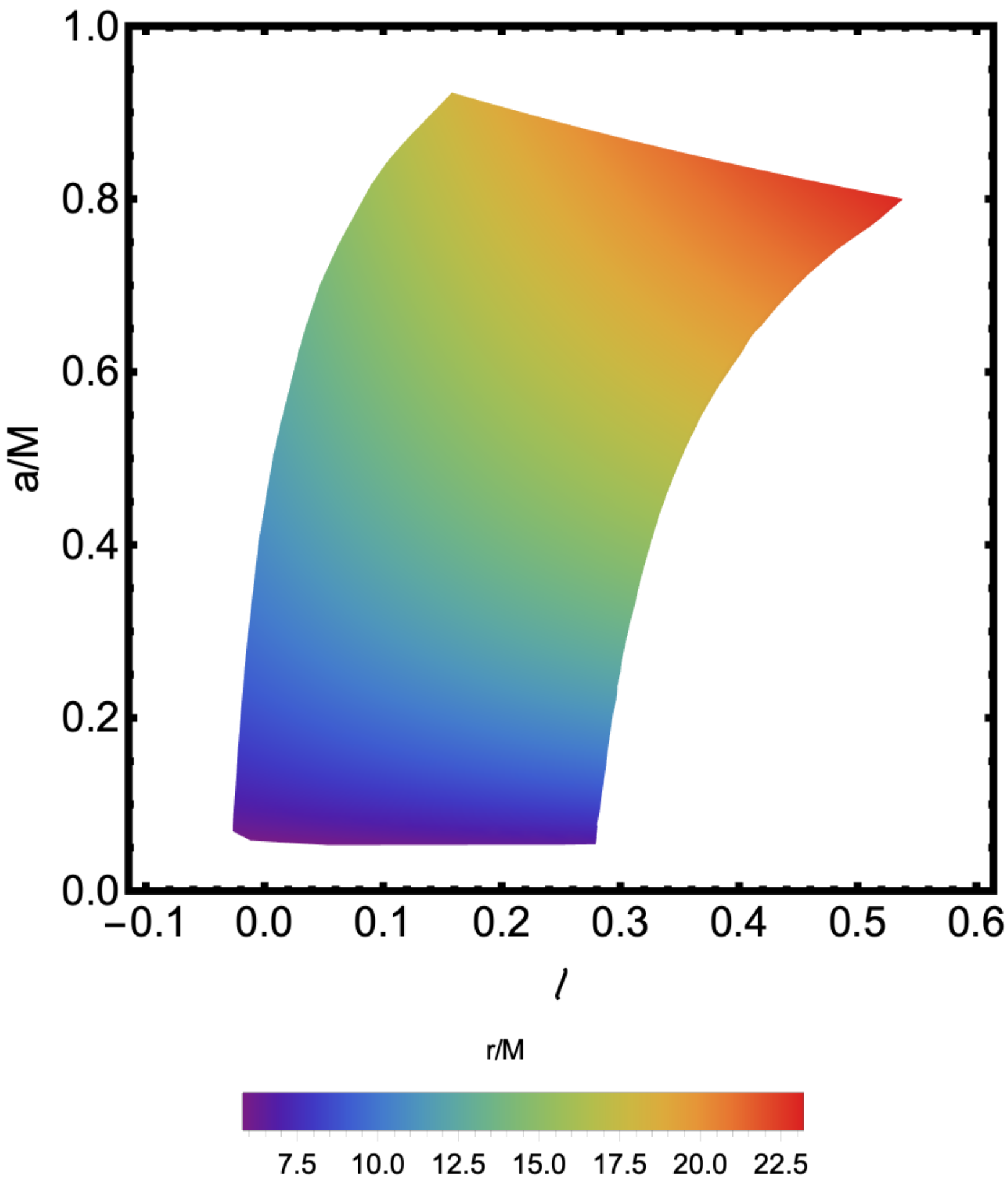}\label{mptlimit}}
\subfigure[retrograde orbit]{\includegraphics[width=0.23\textwidth]{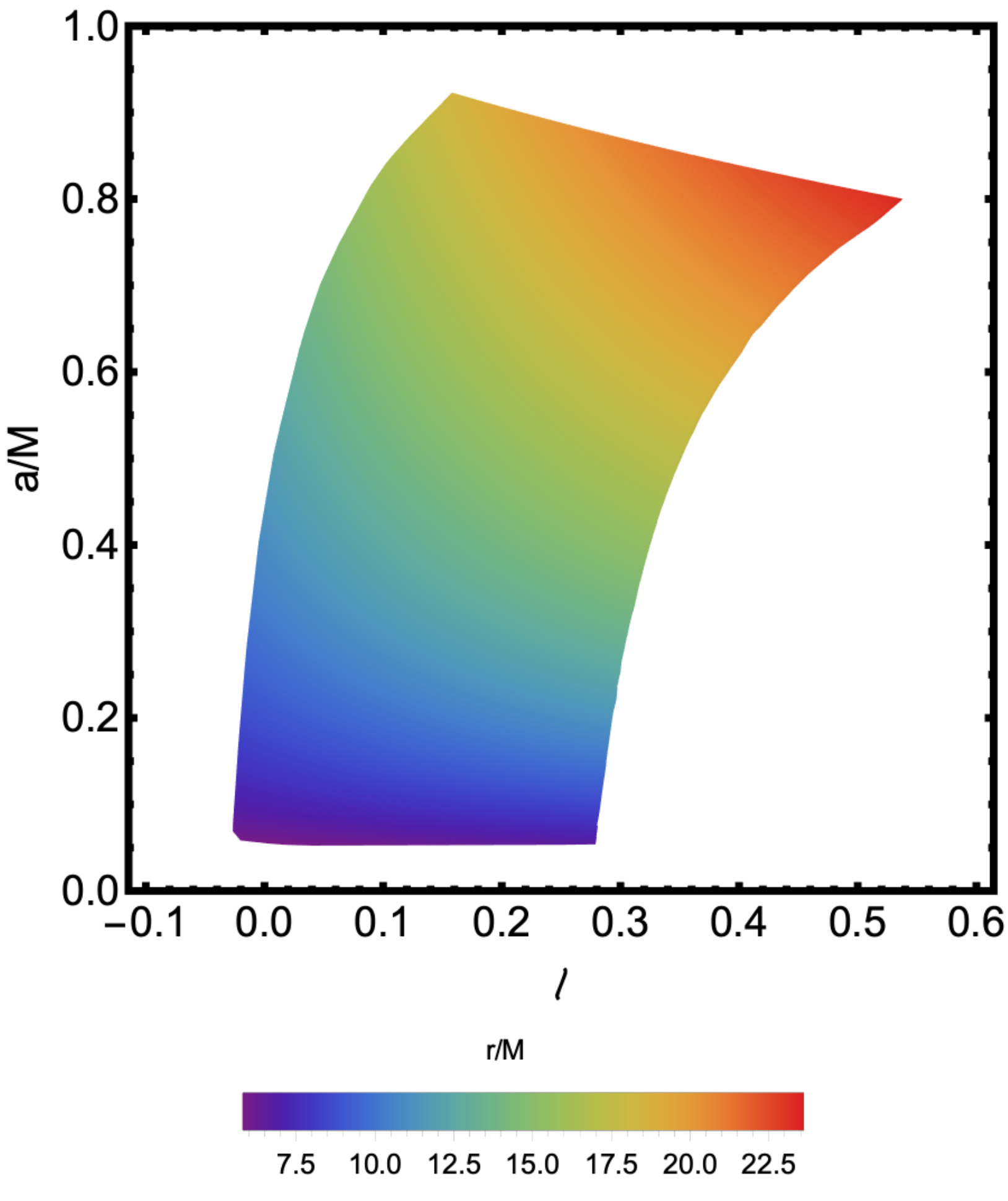}\label{mntlimit}}
\subfigure[]{\includegraphics[width=0.23\textwidth]{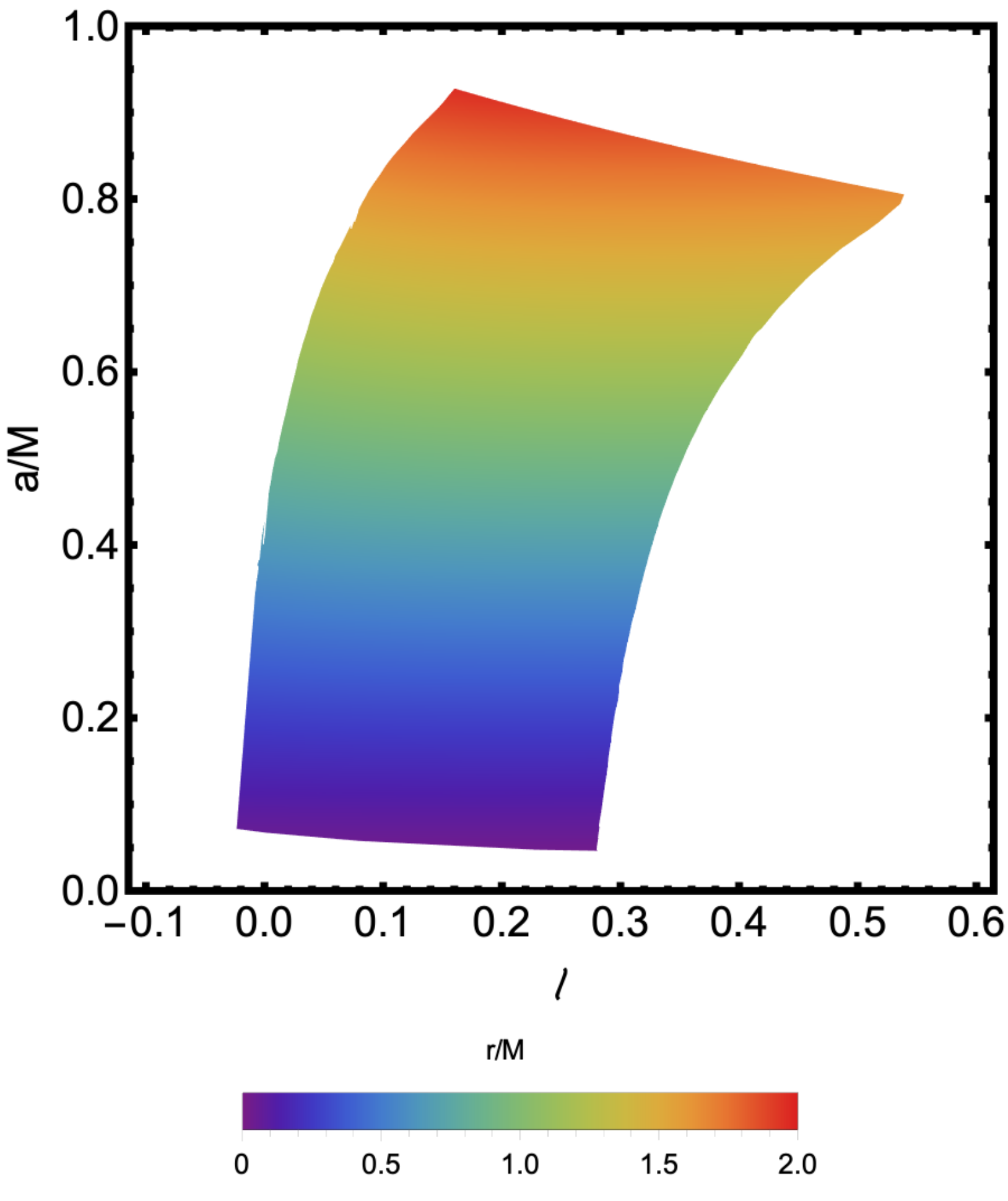}\label{nptlimit}}
\caption{The warp radius $r/M$ of the jet nozzle of M87$^{*}$ constraint with $T=11.24$ year. The range of parameters ($a,\ell$) based on with the angular radius of the shadow and resulted in Ref. \cite{Islam:2024sph}.}\label{figshadow}
\end{figure}

Furthermore, based on the conclusions drawn from the EHT team that the angular radius of the shadow $\theta_{sh}=42\pm3\mu as$, Ref. \cite{Islam:2024sph} provides the corresponding constrained ranges for the spin parameter $a$ and the LSB parameter $\ell$, as shown in Fig. (10). From their results, it can be seen that the angular radius of the shadow give a relatively strong range limit on the spin parameter $a$ and the LSB parameter $\ell$. Then, in conjunction with the constraint range provided in Ref. \cite{Islam:2024sph} and the precession period of the M87* jet, we can derive the corresponding warp radius $r/M$ of the jet, which is illustrated in Fig. \ref{figshadow}. According to condition $T=11.24$ years, the warp radius $r/M$ is confined to $(5.82,22.61)$ for prograde orbits and $(6.17,24.74)$ for retrograde orbits, the corresponding discrepancy between the two orbits falls within the range from $0.05$ to $1.96$. The results indicate that although the angular shadow radius imposes stringent constraints on the $(a,\ell)$ parameter space, a substantial region remains in which rotating LSB black hole can be distinguished. Moreover, the available parameter range remains sufficient to differentiate between prograde and retrograde orbits. 

\section{Conclusion and discussion}\label{conclusion}
In this paper, we investigate massive particle motion around a rotating black hole in Bumblebee gravity and constrain the parameters using observations of the M87* jet precession and shadow radius. Joint constraints from the jet and shadow provide an effective test of supermassive black hole properties and alternative gravity theories. The observed jet precession indicates an inclined accretion disk, motivating spherical orbits as a simplified model linking disk tilt to spacetime geometry.

To analyze the influence of parameters for the rotating black hole in Bumblebee gravity on the motion characteristics of massive particles, we calculated the spherical orbits and ISSOs to assess parameter effects. For non-equatorial orbits, the conserved quantities $\mathcal{E}$, $\mathcal{L}$, and $\mathcal{K}$ depend on $(r,a,\ell,\zeta)$. In prograde motion, $a$ and $\ell$ suppress dynamics, while $\zeta$ decreases $\mathcal{L}$ and increases $\mathcal{E}$ and $\mathcal{K}$. In retrograde orbits, the effect of $a$ reverses; $\mathcal{L}$ and $\mathcal{K}$ grow with $\ell$, while $\mathcal{E}$ decreases, and $\zeta$ reduces $\mathcal{L}$ and $\mathcal{E}$ but enhances $\mathcal{K}$. These trends extend to ISSOs: the ISSO radius $r_{ISSO}$ decreases with $a$ and $\ell$ and increases with $\zeta$ for prograde orbits, and vice versa for retrograde orbits.

Subsequently, we examined the evolution of the angular coordinates $\theta$ and $\phi$ by selecting suitable initial conditions. Analysis of angular evolution shows $\theta$ oscillates within $(\pi/2-\zeta, \pi/2+\zeta)$, while $\phi$ increases nearly linearly, indicating approximately constant azimuthal velocity. Larger $\zeta$ increases $\theta$ oscillation amplitude and deviations from linearity in $\phi$. We compute the period $T_\theta$ and azimuthal accumulation $\phi(T_\theta)/\pi$. In prograde orbits, $T_\theta$ and $\phi(T_\theta)/\pi$ increase with $a$ and $\ell$, while $\zeta$ shortens $T_\theta$ but increases $\phi(T_\theta)/\pi$; increasing $r/M$ lengthens $T_\theta$ and reduces $\phi(T_\theta)/\pi$. In retrograde orbits, $T_\theta$ and $\phi(T_\theta)/\pi$ decrease with $a$ and increase with $\zeta$ and $r/M$; larger $\ell$ prolongs $T_\theta$ and suppresses $\phi(T_\theta)/\pi$. The precession angular velocity $\omega_t$ increases with $a$ and $\ell$ for both prograde and retrograde orbits, grows with $\zeta$ in prograde motion, and decreases with $\zeta$ in retrograde motion; it weakens with increasing $r/M$. The relatively weak effect of $\zeta$ on $\omega_t$ facilitates parameter constraints from M87* jet observations.

Ultimately, we derived the precession period $T$ from Eq. \eqref{eqperiod} and combined it with observational data on the jet and shadow of the supermassive black hole M87*, to determine the black hole jet's warp radius for given black hole parameters. To streamline the constraints, the tilt angle was fixed with $\zeta=1.25^\circ$ based on prior results. For the M87* jet precession period $T=11.24 \pm 0.47$ years \cite{Cui:2023uyb}, the warp radius $r/M$ for given $(a,\ell)$ spans $(5.73,25.15)$ for prograde and $(6.16,26.46)$ for retrograde orbits, increasing with $a$ or $\ell$. Kerr limits ($r/M=14.12$ prograde, $16.1$ retrograde) suggest $r/M>16$ may indicate a non-vacuum Bumblebee vector field. Differences between prograde and retrograde radii range $0.03$–$1.99$, highlighting the stronger effect of $a$ over $\ell$. Including the EHT shadow constraint $\theta_{sh}=42\pm3 \mu$as \cite{Islam:2024sph} restricts $r/M$ to $(5.82,22.61)$ for prograde and $(6.17,24.74)$ for retrograde orbits, with discrepancies $0.05$–$1.96$. It can be seen that the angular radius of the shadow give a relatively strong limit on the warp radius $r/M$.

This method offers a means to constrain black hole parameters, including limits on charge in certain cases. While simplified and in need of refinement through more detailed simulations, the qualitative results are robust and provide a basis for future precision studies. Advances in multi-messenger astronomy may further enhance such constraints through combined observational data.

\begin{acknowledgments}
{This work is supported in part by the Scientific Research Foundation for High-level Talents of Anhui University of Science and Technology, Grant No. (2024yjrc164 and YJ20240001)}.
\end{acknowledgments}

\end{document}